\documentclass[fleqn,usenatbib]{mnras}

\usepackage[T1]{fontenc}

\usepackage{ae,aecompl}
\usepackage{hyperref}
\usepackage[]{lineno}
\usepackage{threeparttable}

\DeclareRobustCommand{\VAN}[3]{#2}
\let\VANthebibliography\thebibliography
\def\thebibliography{\DeclareRobustCommand{\VAN}[3]{##3}\VANthebibliography}
\providecommand{\abs}[1]{\lvert#1\rvert}

\usepackage{graphicx}	
\usepackage{amsmath}	
\usepackage{bm}
\usepackage{caption}
\usepackage{subcaption}

\usepackage{cancel}
\usepackage[dvipsnames]{xcolor}

\usepackage{newtxtext,newtxmath}

\title[Low-mass planets falling into gaps]{Low-mass planets falling into gaps with cyclonic vortices}

\author[Raúl O. Chametla et al.]{
Raúl O. Chametla,$^{1}$\thanks{E-mail: raul@sirrah.troja.mff.cuni.cz (ROC)},
F. J. Sánchez-Salcedo,$^{2}$
Mauricio Reyes-Ruiz,$^{3}$
Carlos Carrasco-González$^{4}$
\newauthor 
and Ond\v{r}ej Chrenko$^{1}$
\\
$^{1}$Charles University, Faculty of Mathematics and Physics, Astronomical Institute V Hole$\check{s}$ovi$\check{c}$k\'ac 747/2, 180 00 Prague 8, Czech Republic\\
$^{2}$Instituto de Astronomía, Universidad Nacional Autónoma de México, Ciudad Universitaria, Apt. Postal 70-264, C.P. 04510, Mexico City, Mexico\\
$^{3}$Instituto de Astronomía, Universidad Nacional Autónoma de México, Ensenada, 22800 B.C, Mexico\\
$^{4}$Instituto de Radioastronomía y Astrofísica (IRyA), Universidad Nacional Autónoma de México (UNAM), Morel\'{i}a Mexico.
}

\date{Accepted XXX. Received YYY; in original form ZZZ}

\pubyear{2015}

\begin{document}
\label{firstpage}
\pagerange{\pageref{firstpage}--\pageref{lastpage}}
\maketitle

\begin{abstract}
We investigate the planetary migration of low-mass planets ($M_p\in[1,15]M_\oplus$, here $M_\oplus$ is the Earth mass) in a gaseous disc containing a previously formed gap. We perform high-resolution 3D simulations with the FARGO3D code. To create the gap in the surface density of the disc, we use a radial viscosity profile with a bump, which is maintained during the entire simulation time. We find that when the gap is sufficiently deep, the spiral waves excited by the planet trigger the Rossby wave instability, forming cyclonic (underdense) vortices at the edges of the gap. When the planet approaches the gap, it interacts with the vortices, which produce a complex flow structure around the planet. Remarkably, we find a widening of the horseshoe region of the planet produced by the vortex at the outer edge of the gap, which depending on the mass of the planet differs by at least a factor of two with respect to the standard horseshoe width.
This inevitably leads to an increase in the corotation torque on the planet and produces an efficient trap to halt its inward migration. In some cases, the planet becomes locked in corotation with the outer vortex. Under this scenario, our results could explain why low-mass planets do not fall towards the central star within the lifetime of the protoplanetary disc. Lastly, the development of these vortices produces an asymmetric temporal evolution of the gap, which could explain the structures observed in some protoplanetary discs.
\end{abstract}




\section{Introduction} \label{sec:intro}

Protoplanetary discs are composed of gas and a small fraction of dust (about 1$\%$ of the total gas density). Remarkably, even with such a low dust density, recent observations of the distribution of dust in several protoplanetary discs around stars of low and intermediate mass (using several techniques and different instruments, e.g: the Atacama Large Millimeter/submillimeter Array (ALMA), Very Large Array (VLA), Keck II and Hubble telescopes) allow us to identify different large scale structures such as: spiral arms \citep{Garufi2013,Grady2013,Benisty2015,Reggiani2018}, gaps \citep{Flock2015, Flock2016}, bright rings \citep{Quanz2013} and large central cavities \citep{Andrews2011, Carrasco2019}. Each of these observed structures can be explained from different theoretical approaches including dust-induced instabilities, secular gravitational instabilities, hydrodynamic or magnetohydrodynamic turbulence, among others (see \cite{Bae2023} and references therein). In addition, gaps and vortices can also be produced by embedded planets at an early stage of their formation \citep{Kep2018,Muller2018,Pinte2018,Pinte2019}.

These structures could play a crucial role in the formation and evolution of planets. In particular, positive radial gradients in the surface density (or in vortensity) may lead to the formation of migration traps \citep[e.g.][]{Masset_etal2006,Bitsch_etal2014,Roma2019,Chrenko2022}. Density changes may  occur in the edges of the dead zones, at the magnetospheric boundary, at the dust sublimation radius, or at the snowline radius \citep{HY2011}.

Local bumps of high density can be formed in the edges of the dead zones because of the differential mass accretion rate between the active zones and the dead zone \citep[e.g.][]{VT2006,Regaly2013}. A weak gradient in the Ohmic resistivity may lead to a transition in the accretion flow rate \citep{Dzyurkevich_etal2010,Lyra_Mac2012,Lyra_etal2015}.
These local density maxima can trigger the formation of
anticyclonic vortices as a consequence of the Rossby-wave instability, which are efficient dust traps \citep[e.g.][]{Lyra2008}.

Using hydrodynamical simulations, \citet{Ataiee_etal2014} study the 
interaction of a planet with a stationary massive anticyclonic vortex created by a density bump.
Initially, the planet migrates towards the bump. Later on, the planet interacts with the
vortex and becomes locked to it.
\citet{Faure_etal2016} investigate the interaction of planets with migrating vortices
in the inner edge of a dead zone. Interestingly, intermediate mass planets remain
trapped but low-mass planets may eventually escape and continue their inward migration.
\citet{Chametla_Chrenko2022} study the impact of vortices formed after destabilization of two pressure bumps on planetary migration. They find that the vortex-induced spiral waves may slow down or even halt the migration of the planets.

In this work we study the effect of a large scale vortex formation at the edges of a surface density gap resulting from an MRI active zone in the outer parts of a protoplanetary disc \citep[e.g.,][]{Flock2015}. Due to the computational cost of 3D resistive magnetohydrodynamic models, we consider purely hydrodynamic 3D models and include a bump in the viscosity of the gas disc to generate a gap in the density profile. Note that unlike the density traps in the inner cavity of protoplanetary discs as previously studied \citep[see for instance][and references therein]{Roma2019}, the transitions in density generated in our gap can arise at different radial distances, not necessarily in the inner disc.

\begin{table}
   \centering
    \caption{Initial conditions and main parameters of our simulations ($r_0$ denotes a reference radius$^*$).}
    \begin{tabular}{lll}
            \hline
            \hline
        & Parameter & Value [code units]\\
            \hline\\
	    \hline
        Mass of the star & $M_\star$ & 1.0 \\
        Gravitational constant  & $G$ & 1.0 \\
        \hline\\
        & Disc &    \\    
         \hline
        Aspect ratio at $r_0$ & $h_0$ & 0.05  \\
        Surface density at $r_0$ & $\Sigma_0$ & $6.366\times10^{-4}$ \\
        Equatorial density slope & $\xi$ & 2.0  \\
        Flaring index & $f$ & 0.5  \\
        $\alpha$-viscosity & $\alpha_0$ &$10^{-3}$ \\
        \hline\\
		& Gap  &\\		
         \hline
	   Inner transition & $r_1$ &0.9 \\
        Outer transition & $r_2$ &1.1 \\
        Transitions width & $s$ &0.05 \\
         \hline\\
		& Planet  &\\		
	   \hline
        Planet mass interval & $M_p$ & $[3\times10^{-6},4.5\times 10^{-5}]$  \\
        Initial planet location & $(r_p,\phi_p,\theta_p)$ & $(1.25,0,0)$ \\
        Softening length & $\epsilon$  & $0.1H(r_p)$\\
        \hline\\
		& Global mesh &\\		
	   \hline
	   Radial extension & $r$ &$[0.5,2.0]$\\
	   Polar extension & $\theta$ & $[\frac{\pi}{2}-3h,\frac{\pi}{2}]$ \\
        Azimuthal extension & $\phi$ & $[-\pi,\pi]$ \\
         \hline\\
		& Additional parameters  &\\
	\hline
	Orbital period at $r_0$& $T_0$ & $2\pi\Omega^{-1}_0$\\
    \hline  
    \end{tabular}
    \begin{tablenotes}
     \item[*] {We consider $r_0=5.2$ au and $M_\star=1M_\odot$ when scaling back to physical units.}
   \end{tablenotes}
    \label{tab:condinit}
\end{table}


Interestingly, our gaps lead to the formation of cyclonic vortices in their edges. Unlike the elongated and anticyclonic vortices reported for instance in \citet{Ataiee_etal2014}, these cyclonic vortices rotate in the same direction to that
of the whole disc and exhibit lower densities and pressures compared
to their surroundings. 
As a consequence, whereas anticyclonic vortices have the capability of dust trapping
\citep[e.g.,][]{Barge&Sommeria1995},  cyclonic vortices are expected to disperse dust particles.
The objective of this work is to study the role of these vortices in the migration of Earth and super-Earth mass planets ($M_p\in[1,15]M_\oplus$, with $M_\oplus$ the Earth mass) located outside the gap. 
In particular, we analyze whether the density gradient can stop migration.

The paper is laid out as follows. In Section \ref{sec:initial}, we describe the gap disc-planet model, code and numerical setup used in our 3D simulations. In Section \ref{sec:results}, we show the results of our numerical models. We present a discussion in Section \ref{sec:discussion}. Finally, the conclusions are given in Section \ref{sec:conclusions}.

\begin{figure}
\includegraphics[width=0.5\textwidth]{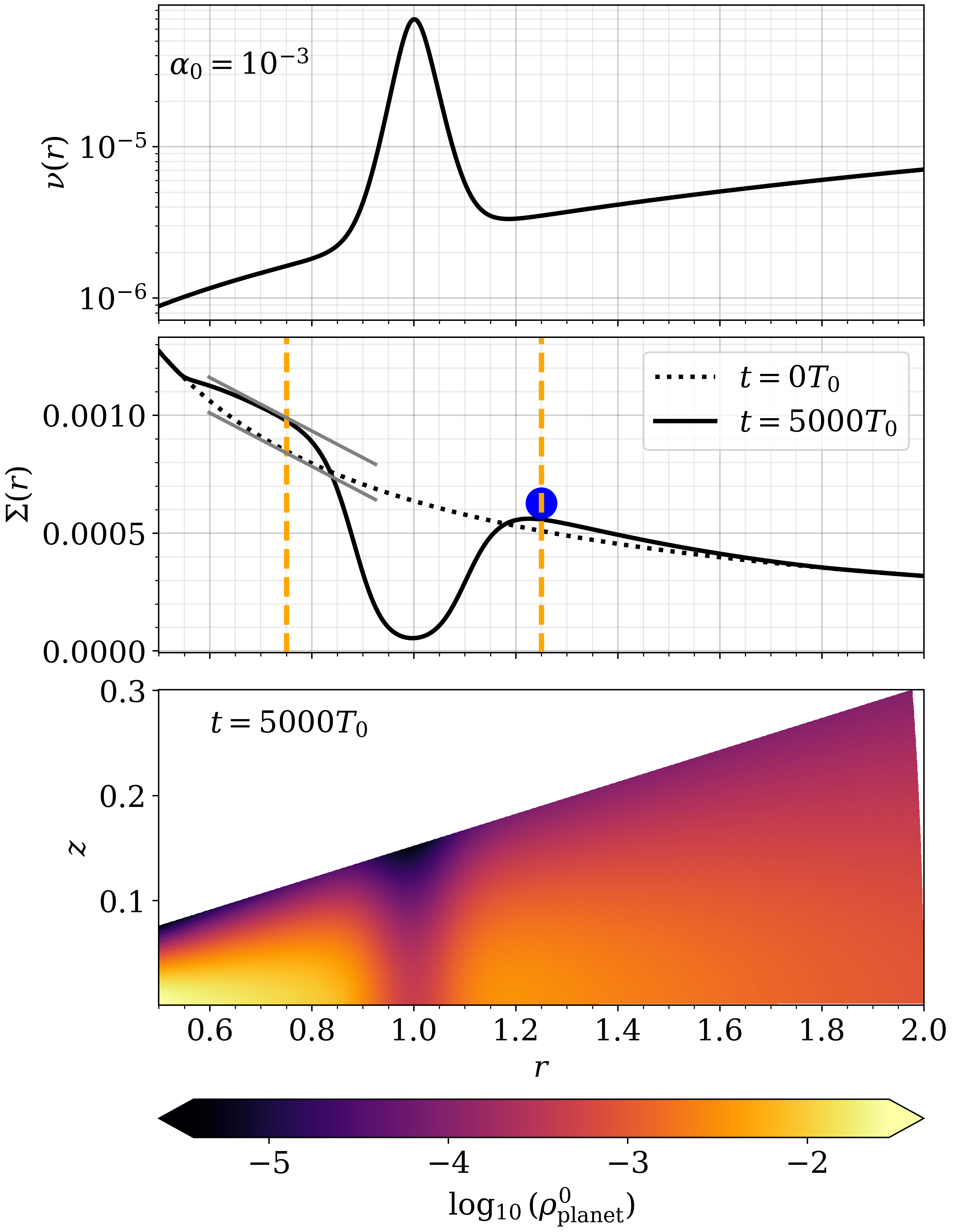}
 \caption{\textit{Top:} Radial profile of the kinematic viscosity $\nu$, parameterized by Eq. (\ref{eq:nu}). \textit{Middle:} Radial profile of the surface density after a time $5000T_{0}$, when
 the disc is viscously relaxed (solid line). The dotted line shows the starting
 power-law surface density profile and the vertical dashed orange lines represent the position of the edges of the gap, which are determined when the slope of the line tangent to the power-law radial profile coincides with the slope of the line tangent to the evolved density profile (see the parallel gray line segments). \textit{Bottom:} Gas density distribution in the $r-z$ plane when the gap profile reaches steady state. Note that at this time the planet is introduced in the disc (that is, $\rho(t=5000T_0)\equiv\rho_\mathrm{planet}^0$).}
\label{fig:2ddens}
\end{figure}

\section{Description of the numerical model}
\label{sec:initial}
In this section, we describe the components of our physical model: the gas disc, the gravitational potential, and we present the code used to solve the set of equations of hydrodynamics.
\subsection{Governing equations}
\label{sec:gas}
We consider a 3D non-self-gravitating gas disc whose evolution is governed by the following equations:

\begin{equation}
\partial_t\rho+\nabla\cdot(\rho \mathbf{v})=0,
 \label{eq:gas_cont}
\end{equation}
\begin{equation}
\partial_t(\rho\mathbf{v})+(\mathbf{v}\cdot\nabla)\mathbf{v}=-\frac{1}{\rho}\nabla p -\nabla\Phi+\mathbf{f}_\nu,
 \label{eq:gas_mom}
\end{equation}
where $\rho$, $\mathbf{v}$ and $\mathbf{f}_\nu$ denote the density, velocity of the gas and the viscous force, respectively. Furthermore, $\Phi$ denotes the gravitational
potential and $p$ is the gas pressure. For the latter, we consider the globally isothermal
equation of state
\begin{equation}
p=c_s^2\rho,
 \label{eq:pressure}
\end{equation}
where $c_s$ is the isothermal sound speed.

The aspect ratio of the disc $h\equiv H/r$, where $H$ is the vertical height of the disc ($H=c_s/\Omega_\mathrm{Kep}$ with $\Omega_\mathrm{Kep}$ is the Keplerian angular frequency) and $r$ the distance to the central star, can be written as:
\begin{equation}
h=h_0\left(\frac{r}{r_0}\right)^f
    \label{eq:aspect_ratio}
\end{equation}
where $f$ is the flaring index and $h_0$ is the aspect ratio at the radial position $r_0$ (see Table \ref{tab:condinit}). In the globally isothermal case $f=0.5$.

We initialize the density and the gas velocity components in a similar way as described in Appendix A in \citet[][]{MB2016} for a globally isothermal disc:
\begin{equation}
\rho=\rho_\mathrm{eq}(\sin\theta)^{-\xi}\exp{\left[h^{-2}\left(1-\frac{1}{\sin\theta}\right)\right]},
    \label{eq:rho}
\end{equation}
where $\theta$ is the polar angle and
\begin{equation}
\rho_\mathrm{eq}=\frac{\Sigma_0}{\sqrt{2\pi}h_{0}r_0}\left(\frac{r}{r_0}\right)^{-\xi},
    \label{eq:rhoeq}
\end{equation}
with $\Sigma_0$ is the surface density\footnote{The surface density and the midplane volumetric density can be related by $\Sigma(r)=\sqrt{2\pi}\rho_\mathrm{eq}H$.} at $r=r_0$.

For the velocity components we assume that $v_r=v_\theta=0$ and 
\begin{equation}
v_\phi=\sqrt{\frac{GM_\star}{r\sin{\theta}}-\xi c_s^2},
 \label{eq:vphi}
\end{equation}
where $M_{\star}$ is the mass of the central star.

The gravitational potential $\Phi$ is given by
\begin{equation}
\Phi=\Phi_\star+\Phi_p,
 \label{eq:potential}
\end{equation}
where
\begin{equation}
\Phi_\star=-\frac{GM_\star}{r},
 \label{eq:Star_potential}
\end{equation}
and
\begin{equation}
\Phi_p=-\frac{GM_p}{\sqrt{r'^2+\varepsilon^2}}+\frac{GM_pr\cos\phi\sin\theta}{r_p^2}
 \label{eq:Planet_potential}
\end{equation}
are the stellar and planetary potentials, respectively. In
Eq.~\eqref{eq:Planet_potential}, $M_p$ is the planet mass, $r'\equiv\abs{\mathbf{r}-\mathbf{r}_p}$ is
the cell-planet distance, $\phi$ is the azimuth with respect to the
planet, and $\varepsilon$ is a
softening length used to avoid computational divergence of the potential in the vicinity of the planet. The second
term on the right-hand side of Eq.~\eqref{eq:Planet_potential} is the
indirect term arising from the reflex motion of the star. Our simulations were
performed with $\varepsilon = 0.1H$, which is comparable to the size of two cells of our numerical mesh (see below). 
We have done some experiments with a larger $\varepsilon$ and found similar results.

\begin{figure}
\includegraphics[width=0.44\textwidth]{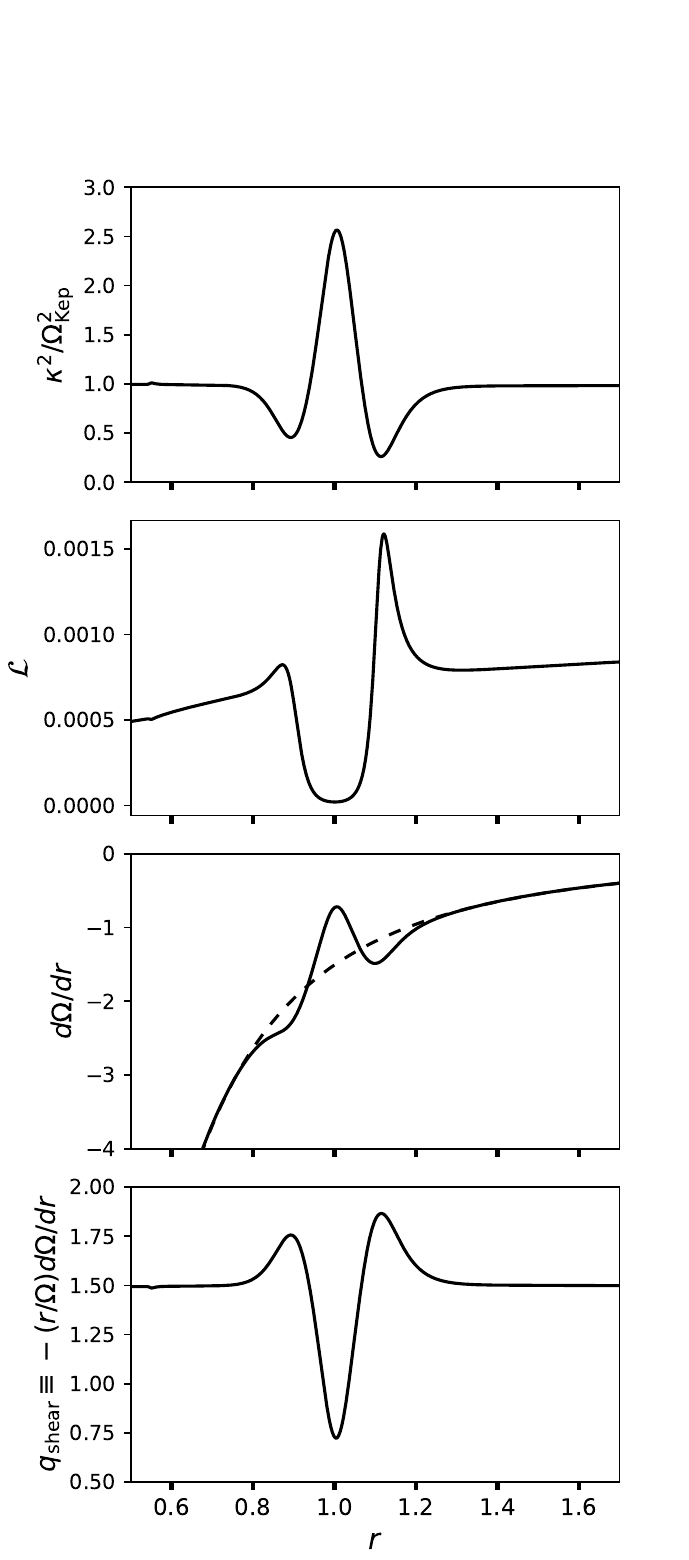}
 \caption{Radial profile of some quantities along the gap before the planet is inserted.
 From top to bottom: (1) squared epicyclic frequency relative to the squared Keplerian frequency,
 (2) inverse of the potential vorticity
 $\mathcal{L}$, (3) $d\Omega/dr$ as a function of radius and (4) shear parameter $q_\mathrm{shear}\equiv -(r/\Omega)d\Omega/dr$.}
\label{fig:stability}
\end{figure}

\begin{figure*}
\includegraphics[scale=0.53]{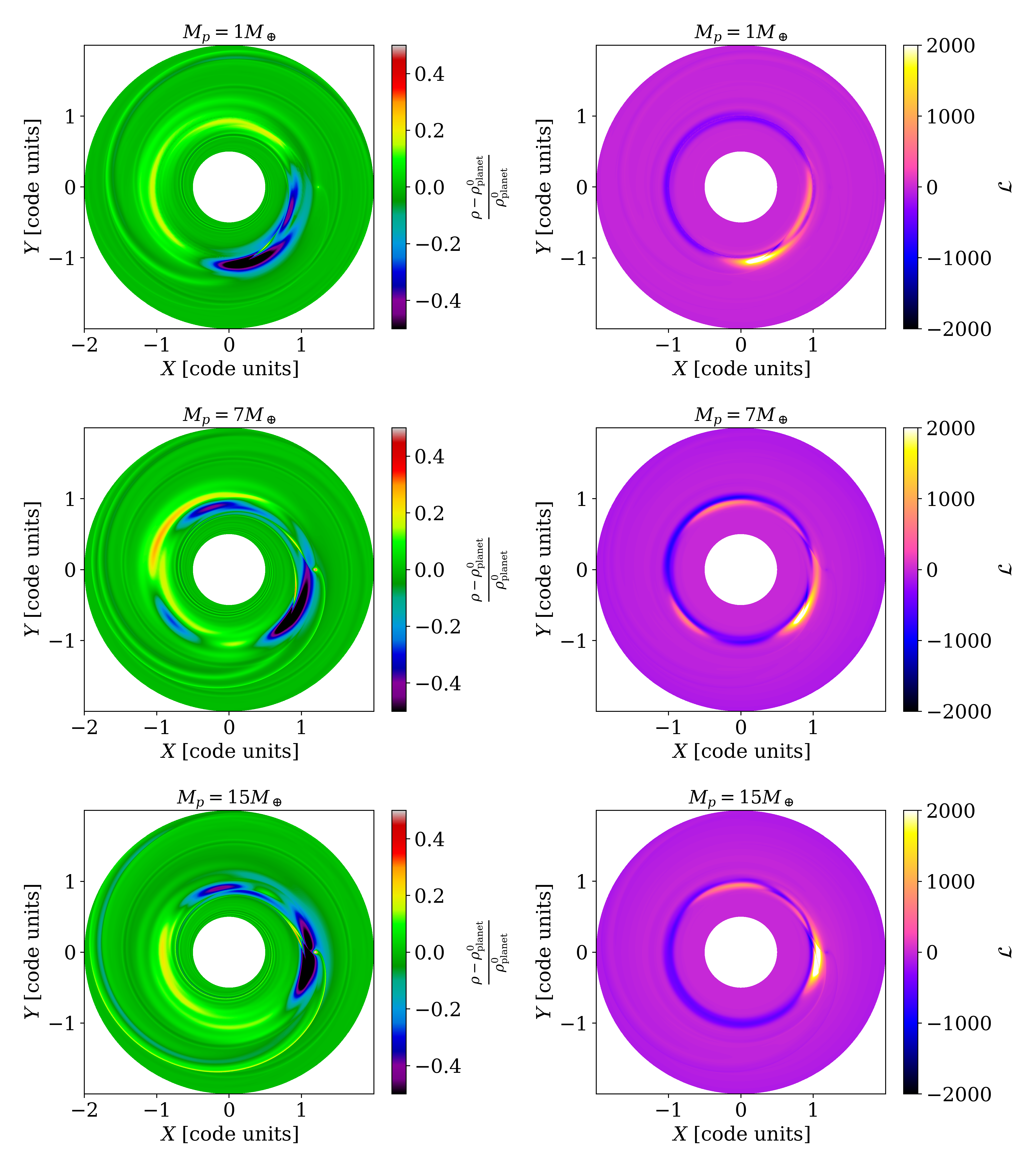}
  \caption{\textit{Left:} Gas density perturbation at the midplane for three different planetary masses  ($M_p=1,7,15M_\oplus$) at $t=500T_0$.  \textit{Right:} Residual vortensity for the same planetary masses also at $t=500 T_0$.}
\label{fig:DV-maps}
\end{figure*}

\begin{figure}
\includegraphics[width=0.5\textwidth]{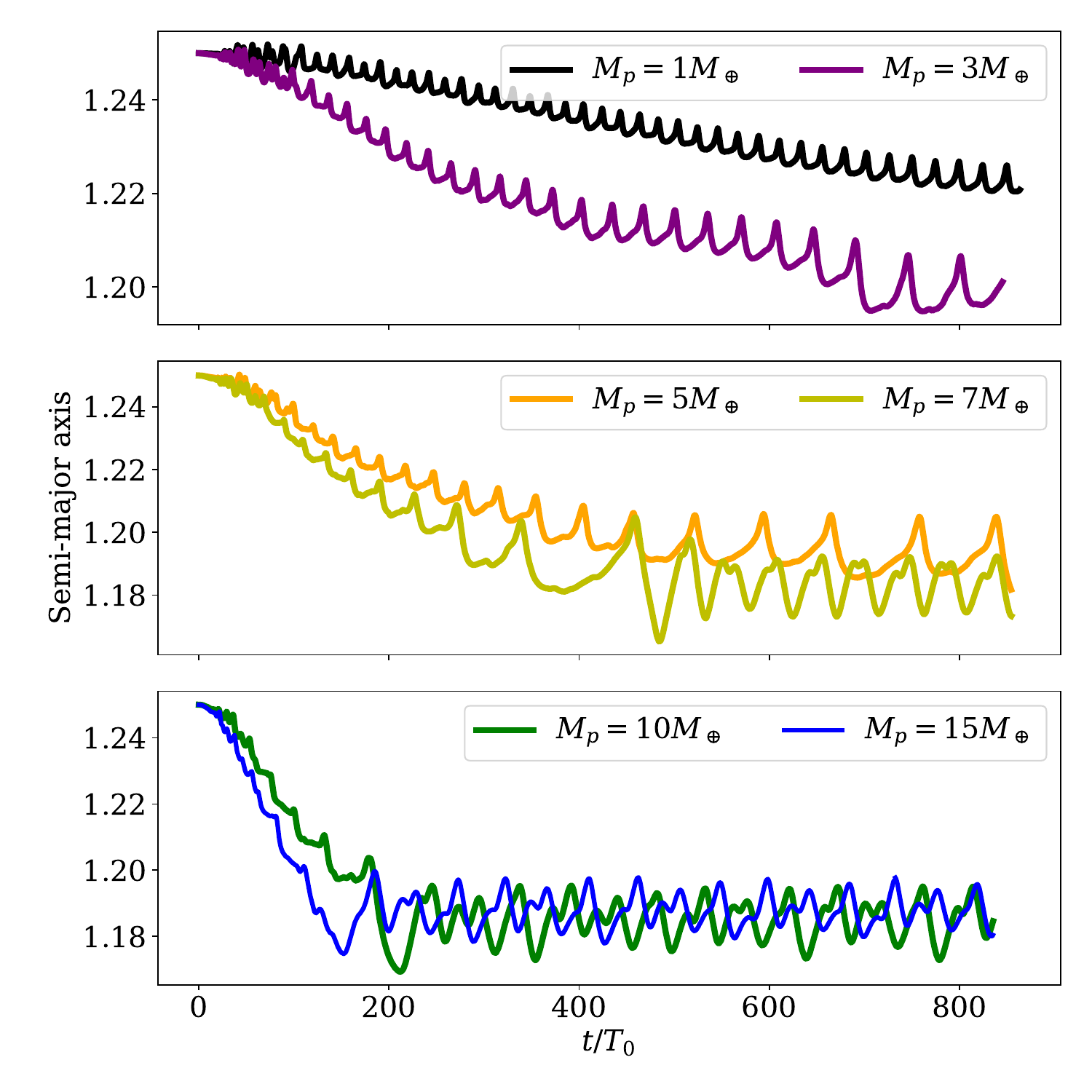}
  \caption{Temporal evolution of the semi-major axis for different planetary masses.}
\label{fig:semi}
\end{figure}

\begin{figure*}
\includegraphics[scale=0.5]{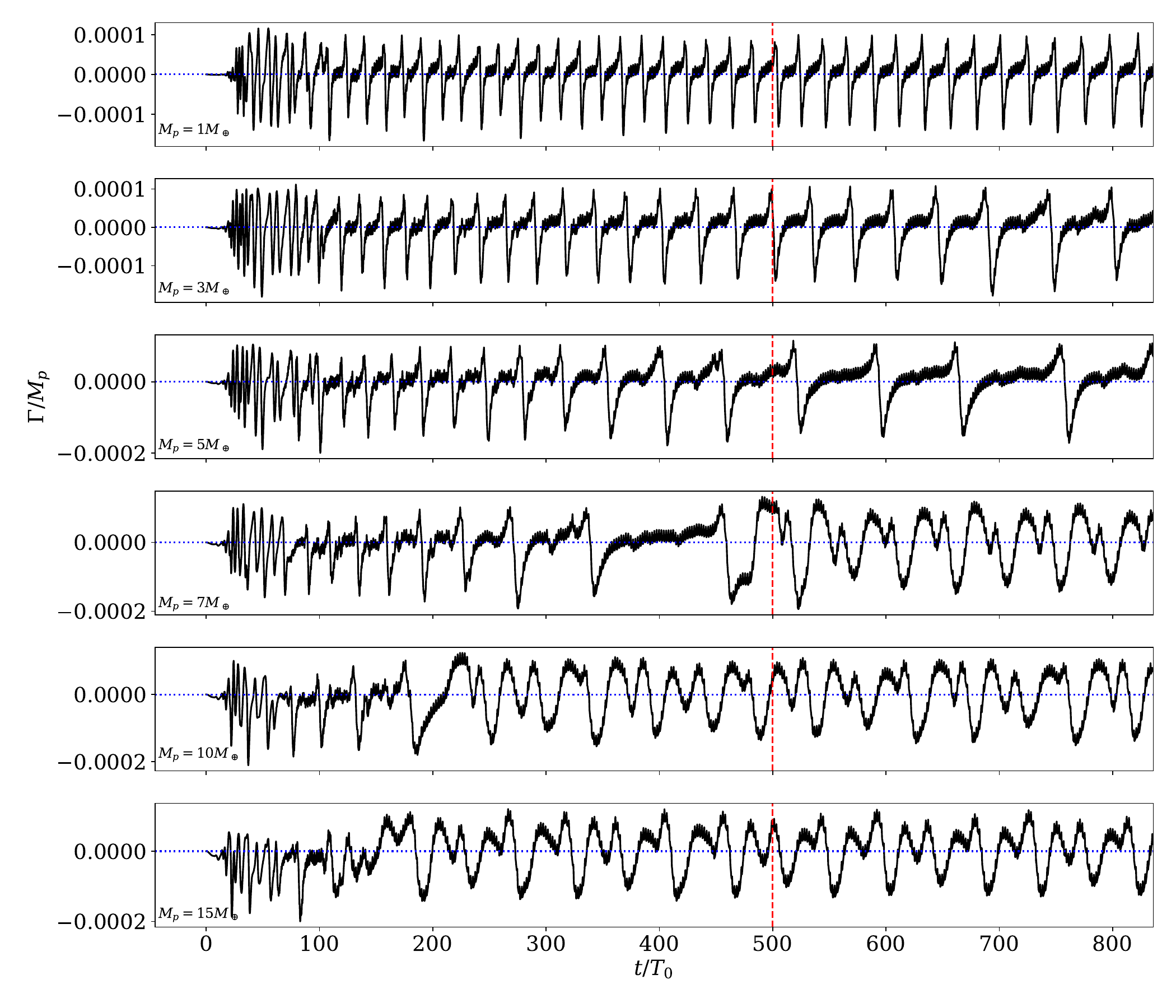}
  \caption{Temporal evolution of the total specific torque on the planets. The vertical red dashed lines indicate $t=500T_{0}$,
  which corresponds to the time at which the snapshots of Figure \ref{fig:DV-maps} were taken.}
\label{fig:Tdt}
\end{figure*}

\begin{figure*}
\includegraphics[scale=0.55]{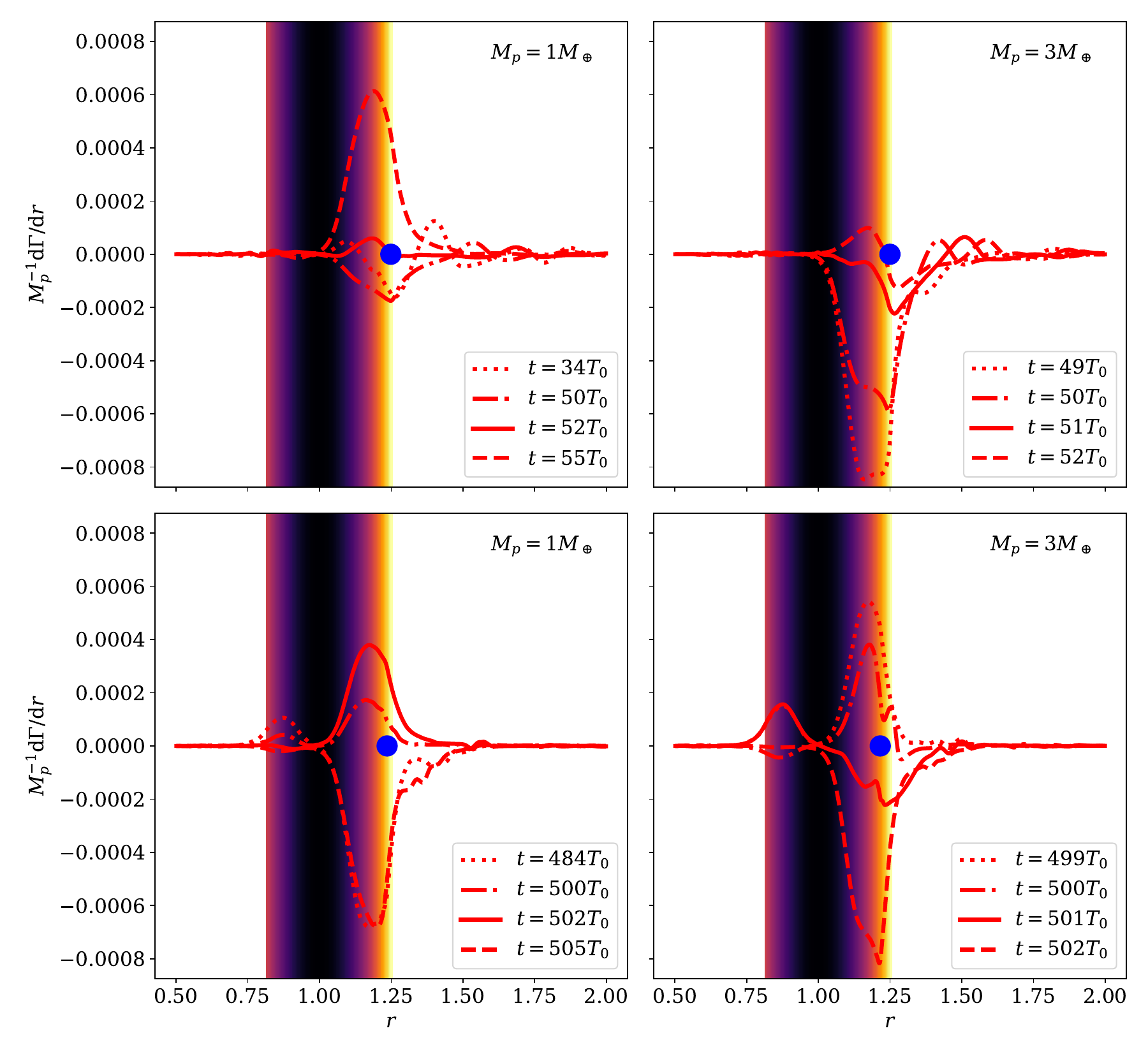}
  \caption{Radial torque distribution (in code units) for planetary masses of $M_p=1M_\oplus$, $M_p=3M_\oplus$ at different times, including those times at which the torque reaches its maximum and minimum values. The blue dot shows the radial position of the planet, and the colored area denotes the relative radial extent between the inner and outer edges of the gap.}
\label{fig:radial_torq}
\end{figure*}

\begin{figure}
\includegraphics[width=8cm,height=14cm]{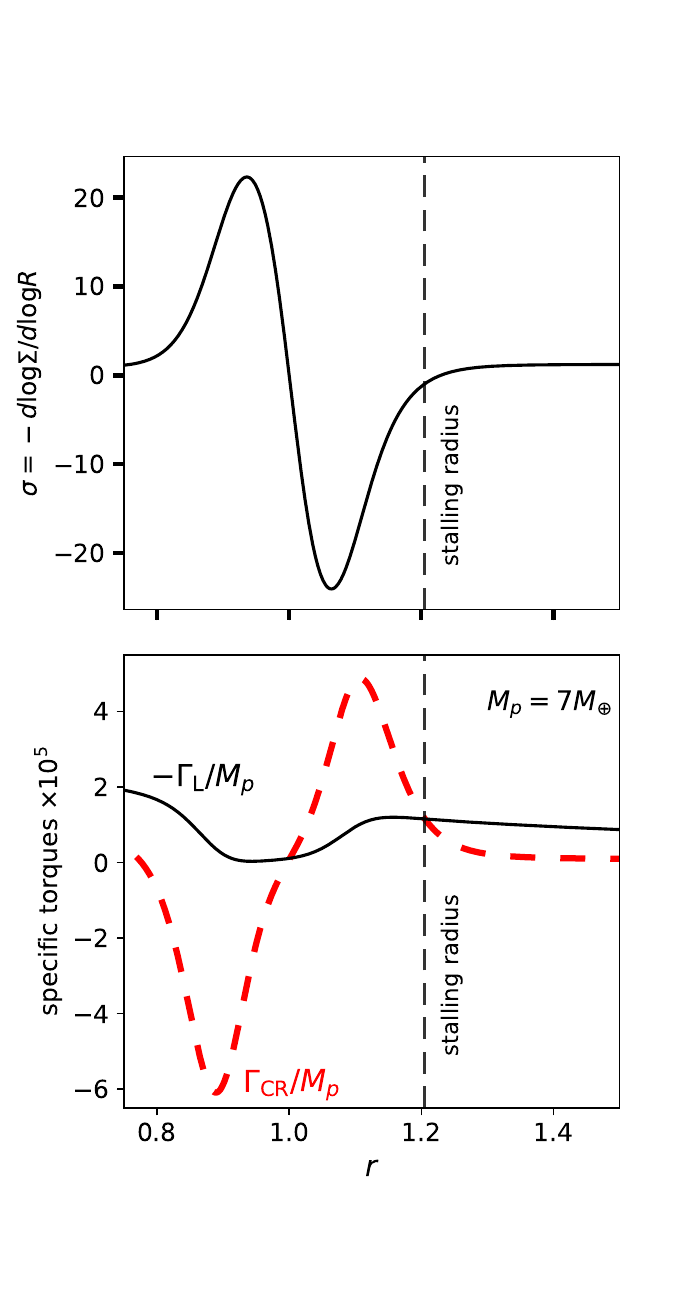}
 \caption{Radial profile of $\sigma\equiv -d\ln\Sigma/d\ln r$ (top panel), and the specific Lindblad and corotation torques
 for $M_{p}=7M_{\oplus}$ using Eqs. (\ref{eq:tanaka})-(\ref{eq:corotation_tq_formula}) (bottom panel). For the half-width
 of the horseshoe region $x_{s}$, we have used the formula provided in \citet{JM2017}.
 The vertical dashed lines indicate the radius at which $\Gamma_{\rm L}+\Gamma_{\rm CR}=0$ (stalling radius).}
\label{fig:CR_torque}
\end{figure}

\begin{figure*}
\includegraphics[scale=0.57]{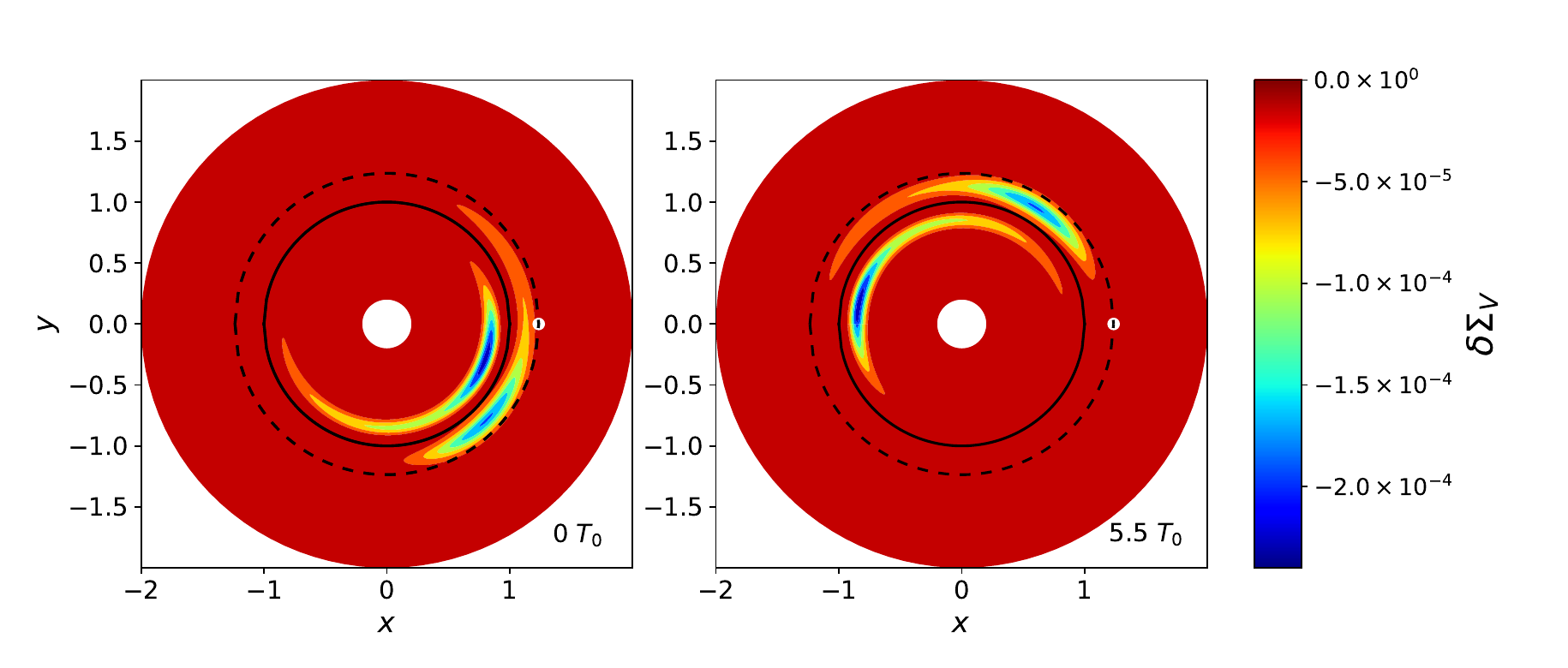}
\caption{Non-axisymmetric decrement in the surface density, $\delta\Sigma_{V}$ (in code units), 
representing two underdense vortices, at $t=0$ (left) and at $t=5.5 T_{0}$ (right).
The two vortices and the planet are rotating counterclockwise around the central object.
The frame is corotating with the planet (white dot). In this illustrative example, it was assumed that
the planet is fixed at $r=1.236$. For reference, a unit circle 
has been drawn (solid line).
The dashed circle indicates the planetary orbital radius.
\label{fig:Sigma_simple_model}}
\end{figure*}

\begin{figure*}
\includegraphics[width=19cm,height=9cm]{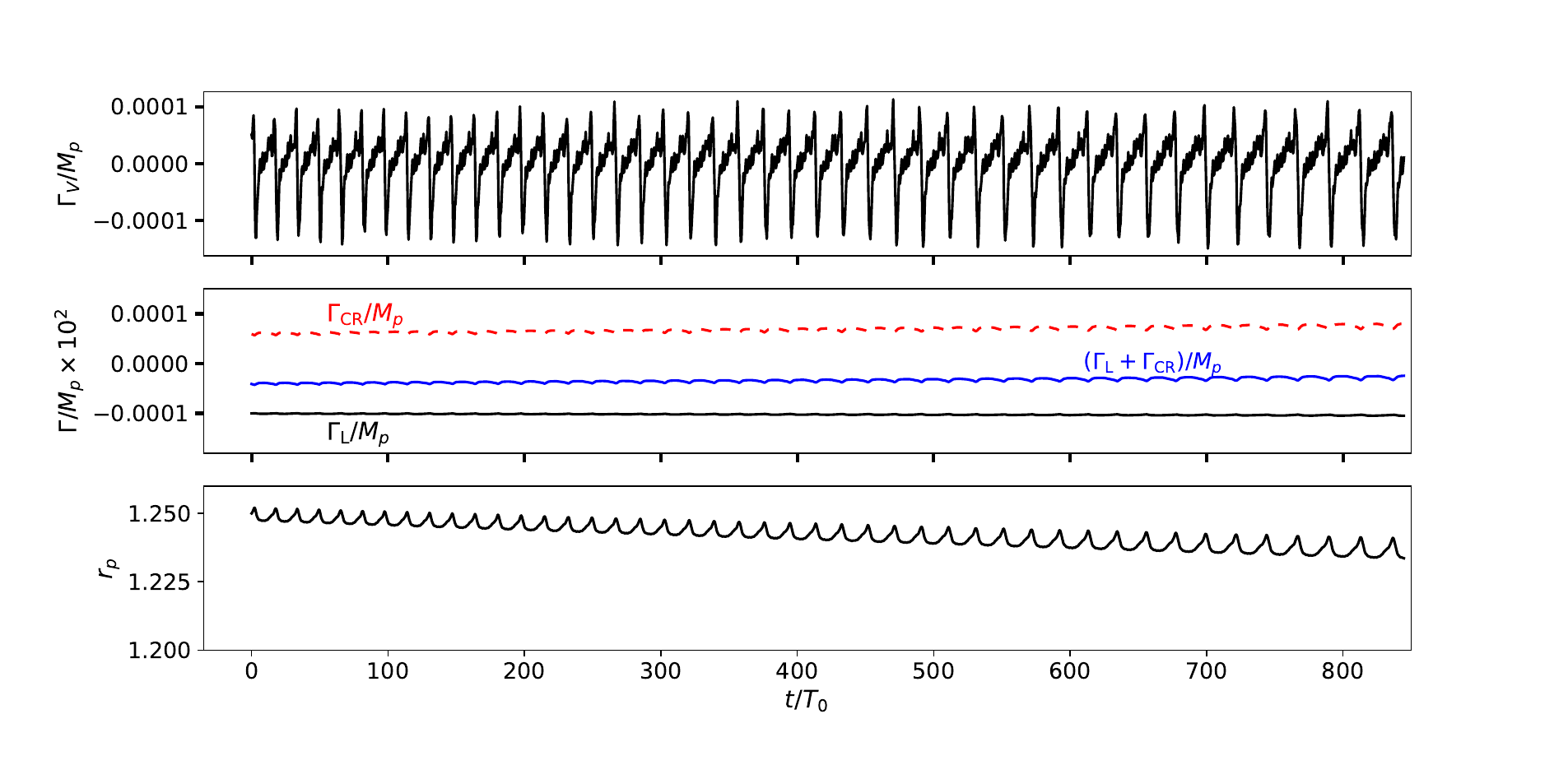}
\caption{Temporal evolution of the specific torques and planetary orbital radius in the semi-analytical model
described in Section \ref{sec:simple_model}, for a $1M_{\oplus}$ planet initially at a radius of $1.25$. 
Upper panel: $\Gamma_{V}/M_{p}$; middle panel: $\Gamma_{\rm L}/M_{p}$ and
$\Gamma_{\rm CR}/M_{p}$; lower panel: $r_{p}(t)$.}
\label{fig:tq_rp_simple_model_1Moplus}
\end{figure*}

\begin{figure*}
\includegraphics[width=19cm,height=9cm]{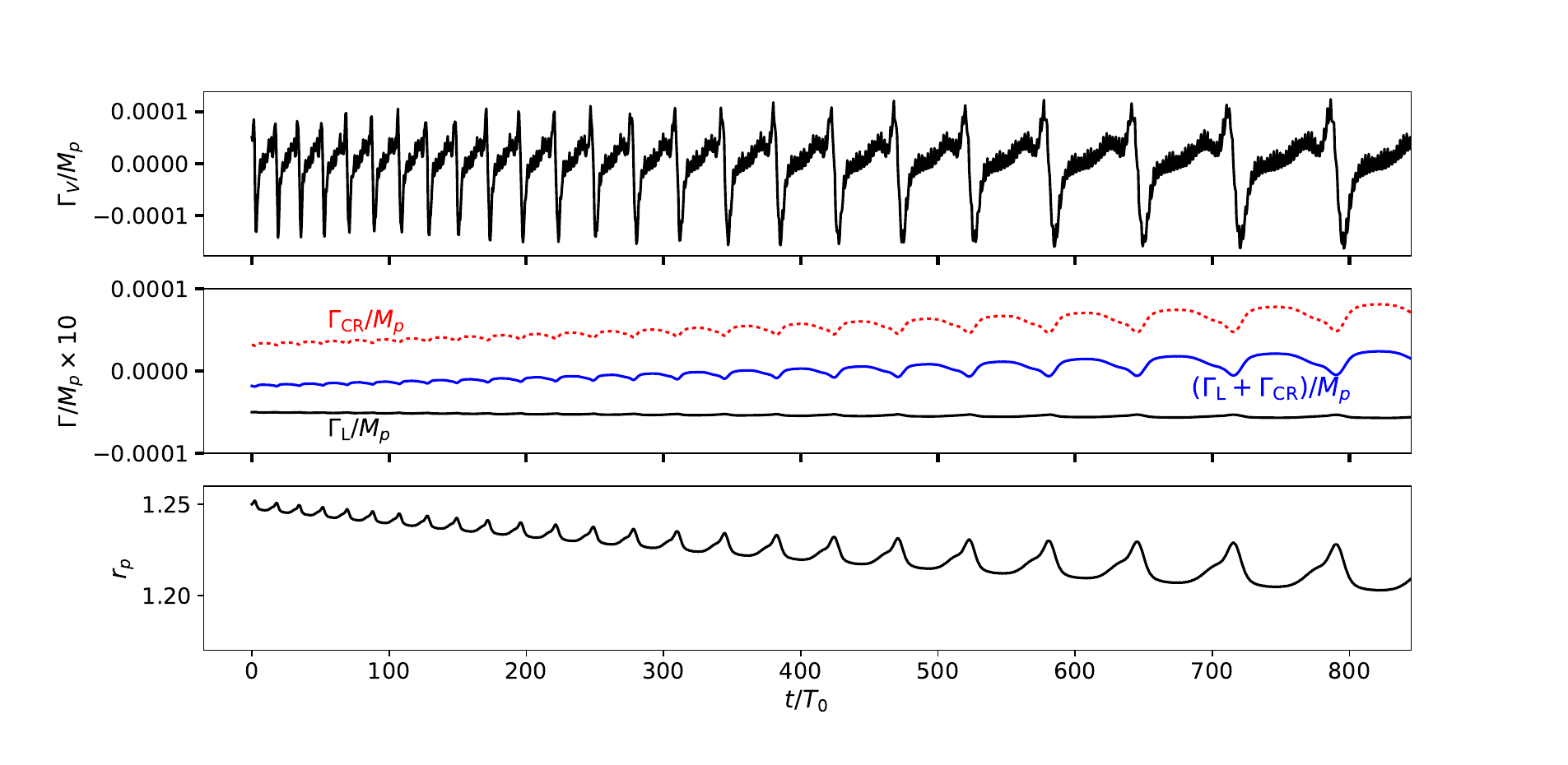}
\caption{Same as Figure \ref{fig:tq_rp_simple_model_1Moplus} but for $M_{p}=5M_{\oplus}$.}
\label{fig:tq_rp_simple_model_5Moplus}
\end{figure*}

\subsection{Code and set-up} \label{sec:model}
We use the publicly available hydrodynamic code {FARGO3D\footnote{
https://fargo3d.github.io/documentation}
\citep[][]{Benitez_Masset2016} which includes the fast orbital advection algorithm \citep[see][for details]{Masset2000,Benitez_Masset2016}, which significantly increases the integration timestep in numerical studies
of protoplanetary discs. We use a computational mesh grid with 
uniform spacing in radius, azimuthal and polar directions. We simulate the disc over the
radial range $r\in[0.5,2.0]r_0$, an azimuthal extent of $\phi=[-\pi,\pi]$ and a polar extent of $\theta=[\pi/2-3h_0,\pi/2]$. The number of grid cells is $(N_r,N_\phi,N_\theta)=(768,3200,76)$. For our adopted value of $h_{0}=0.05$, this corresponds to a grid size
of $1.97 \times 10^{-3}r_{0}$ in each direction or, equivalently, $26$ cells per pressure length-scale.
This resolution is similar to that used in \citet{MB2016}, where the dynamics of the 3D horseshoe region is studied. 

Particular attention was paid to avoid reflections of the spiral waves excited by the planet in the radial boundaries.
To this end, we use damping boundary conditions as in \citet{deVal2006}. The width
of the inner damping ring is $3.9\times 10^{-2}r_0$ and that of the
outer ring being $8.54\times 10^{-1}r_0$. The damping timescale at the edge of each damping ring equals $1/20^{th}$ of the local orbital period.
Since we only model one hemisphere of the disc, we use reflecting boundary conditions at the midplane. At the upper boundary of the disc, the gas density and azimuthal velocity component are extrapolated from the initial conditions, whereas for the radial and polar velocity components we apply reflecting boundary conditions.

In Table~\ref{tab:condinit} we present the set of parameters used in our numerical models.

\subsection{Gap modeling through radial viscosity transitions}

To make a gap in the surface density of the disc, we set up an
axisymmetric viscosity bump in a ring around $r_0$.  The viscosity bump, which is kept
constant with time, is given by the function:
\begin{equation}
    \nu(r) =\alpha c_s H,
    \label{eq:nu}
\end{equation}
with
\begin{equation}
\alpha(r) =\alpha_0\left\{1-\frac{1}{2}\left[\tanh\left(\frac{r-r_1}{s}\right)-\tanh\left(\frac{r-r_2}{s}\right)\right]\right\}^{-1}.
\label{eq:alpha}
\end{equation}
In Eq. (\ref{eq:alpha}), $\alpha_0$ is the viscosity parameter outside the bump,
$r_1$ and $r_2$ represent the radial positions of the inner/outer viscosity transitions 
in the gap edges, and $s$ is the width of these transitions. The viscosity bump
has a width $\Delta r=r_{2}-r_{1}$ and it takes a peak value, $\alpha_{\rm max}$,
given by
\begin{equation}
\alpha_{\rm max}=\frac{\alpha_{0}} {1-\tanh\left(\frac{\Delta r}{2s}\right)},
\end{equation}
at a radius $r=(r_{1}+r_{2})/2$. Hence, the fractional change of $\alpha$ across the bump,
$\alpha_{\rm max}/\alpha_{0}$, decreases with $s$
if $\Delta r$ is kept fixed.

We start with a disc with a power-law surface density with a slope given by $\sigma \equiv -d\ln\Sigma/d\ln r=1$, and then we bring the disc to equilibrium by performing axisymmetric $(r,z)$ runs without including the planet's gravitational potential.
Figure \ref{fig:2ddens} shows the viscosity and the relaxed surface density at $5000T_{0}$,
as a function of $r$, for $\alpha_0=10^{-3}$, $r_{1}=0.9$, $r_{2}=1.1$ and $s=0.05$.
The density map in the $r-z$ plane is also shown.
We see that the gap does not have pronounced bumps at its edges, similar to the gap formed in a turbulent 
disc \citep[see, for instance, Fig. 2 of the D2G$\_$e-2 model in][]{Flock2015}. 
Once we know the volume density in the $r-z$ plane, we 
expand the grid into the azimuthal direction and the planet is included.

For clarity of presentation, we will focus on simulations with the values of 
$\alpha_{0}$, $r_{1}$, $r_{2}$, and $s$, as quoted above,
throughout the main body of the paper. However, the results for a shallower gap 
(a larger value of $s$) are described in the Appendix. We also show in the Appendix that a reduction in the height of the bump
(ie., a lower value of $\alpha_{\rm max}/\alpha_{0}$) can generate a gap  where vortex formation is not
viable.

The adopted value for the background viscosity, $\alpha_{0}=10^{-3}$, 
falls within the upper range of the observed values 
\citep[][]{Pinte_2016,Flaherty_2020,Jiang_2024}. Nevertheless, 
when adopting a smaller value of $\alpha_{0}$ (keeping
$r_{1}$, $r_{2}$ and $s$ fixed), the relaxed surface density
profile remains essentially unaltered, as the new $\alpha(r)$ is just a rescaled profile
\citep[e.g.,][]{LP1974}. However, a reduction of $\alpha_{0}$ by a factor of, say, $10$
implies that the stationary surface density is reached in a timescale $10$ times
longer. To avoid this additional computational cost, we chose $\alpha_{0}=10^{-3}$.
If vortices survive for $\alpha_{0}=10^{-3}$, it is likely that they will also survive
for smaller values of $\alpha_{0}$ because their lifetimes typically increase with lower 
values of $\alpha$ \citep[e.g.,][]{Regaly2017}.

\subsection{Stability analysis}
Pressure bumps and gaps can be unstable to axisymmetric perturbations. Since our disc is globally isothermal (barotropic),
the Rayleigh condition for local axisymmetric stability is $\kappa^{2}\geq 0$, where $\kappa$ is the epicyclic frequency.
Note that the angular velocity of the disc $\Omega$ differs from $\Omega_{\rm Kep}$ because of the thermal pressure gradient.
The top panel of Figure \ref{fig:stability} shows that the minimum value of $\kappa^{2}$ is $0.26\Omega_{\rm Kep}^{2}$ and, therefore, the
gap is stable to axisymmetric perturbations. Additionally, we mention that although the gas flow is not completely Keplerian in the disc, it is dynamically stable except at the edges of the gap where the shear parameter $q_\mathrm{shear}\to2$ (see lower panel in Fig. \ref{fig:stability}) and nonlinear instabilities can arise \citep{Hawley1999}.

On the other hand, Rossby wave instability (RWI) may occur (not necessarily) if $\mathcal{L}\equiv \Sigma/(2 \omega_{z})$ has a maximum
\citep{Lovelace_etal1999,Li_etal2000}. Here $\omega_{z}$ is the vertical vorticity $\omega_{z}= (\vec{\nabla}\times \vec{v})_{z}$.
The central panel of Fig. \ref{fig:stability} shows that $\mathcal{L}$ presents two maxima in our disc, one at each edge of the gap.
Thus, our gap is potentially prone to the RWI. Moreover, \citet{Chang2023} find empirically that the RWI takes place if the condition $\kappa^{2}+N_{R}^{2} \lesssim
0.6\Omega_{\rm Kep}^{2}$ is satisfied somewhere in the disc. In our case,
as the Brunt-V\"ais\"al\"a frequency $N_{R}$ is zero, this condition implies
$\kappa^{2}\lesssim 0.6\Omega_{\rm Kep}^{2}$, which is fulfilled in the edges of the
gap (see Fig. \ref{fig:stability}). Hence, we expect the gap to be unstable to the RWI.

\section{Results}
 \label{sec:results}

In this section we present the results of our 3D simulations of a planet migrating in our gapped protoplanetary disc. We are mainly interested in the orbital evolution of the planet and in the torques that drive its migration. 
We introduce the planet on a circular orbit with initial radius $1.25r_0$ and migrates towards the gap.
To avoid any artificial disturbance due to the sudden introduction of the planet, we have introduced the planet using the following mass-taper function:
\begin{equation}
M_p(t)=M_p\left\{
	       \begin{array}{ll}
		 \frac{1}{2}\left[1-\cos\left(\dfrac{\pi t}{t_\mathrm{mt}}\right)\right]      & \mathrm{if}\, t<t_\mathrm{mt}\\
		 1 & \mathrm{otherwise} \\
	       \end{array}
	     \right.
    \label{eq:Mp_t}
\end{equation}
where $t_\mathrm{mt}$ is the time-scale over which the mass of the planet grows to its constant value, which we set to five orbital periods in all our simulations. 

\subsection{Cyclonic vortices}
Fig. \ref{fig:DV-maps} shows the density perturbation in the midplane of the disc, $(\rho-\rho_\mathrm{planet}^0)/\rho_\mathrm{planet}^0$, where $\rho_\mathrm{planet}^0$ 
is the unperturbed volume density, at $5000T_0$ (just when the planet is inserted). We also show the residual
vortensity at the same time. Contrary to classical gaps where the vortices are formed in the pressure maxima of gap edges
\citep[e.g.,][]{Li_etal2000}, we see that vortices are formed within the gap.
It is remarkable that the vortices formed at the edges of gap have a cyclonic circulation. As a consequence, the surface density in the vortex and its immediate vicinity is lower than that of the surrounding regions (see Fig.  \ref{fig:DV-maps}). This gives rise to a novel interaction between the vortices in protoplanetary discs and planetary bodies which is the focus of the present work. It is usually argued that cyclonic vortices in protoplanetary discs are rapidly destroyed by the shear flow \citep[e.g.,][]{GL1999}.
However, \citet{Love2009} envisage the potential formation of cyclonic vortices as a consequence of the RWI in locally non-Keplerian discs in regions where $d\Omega/dr>0$. 
Our simulations indicate that the condition $d\Omega/dr>0$ is not a necessary condition (see bottom panel in Figure \ref{fig:stability}).
The formation of cyclonic vortices is an interesting feature found in our study, and will be analized in detail in a follow-up paper currently in preparation.

\subsection{Orbital evolution}
\label{subsec:oe}

Fig. \ref{fig:semi} shows the semi-major axis versus time, for planets with masses $M_p\in[1,15]M_\oplus$. 
At the end of the simulation,
the two less massive planets are still migrating towards the gap. However, the two most massive planets rapidly migrate 
down to $r=1.18 r_0$, and then halt their inward migration. In fact, they remain within the outer edge of the gap which is located at $r=1.25r_0$ (see Fig. \ref{fig:2ddens}). 

Note that prior and after reaching the stalling radius, the semi-major axes of the planets exhibit oscillations. 
As we will see in Section \ref{sec:simple_model}, these oscillations are due to the interaction of the planet with the vortices formed at the edges 
of the gap (see Fig. \ref{fig:DV-maps}). On the other hand, we mention that for the range of planetary masses studied here, the eccentricity of the planet does not develop considerably. Since the maximum value of the eccentricity that we find is $e_p=5.12\times10^{-4}$ which is damped quickly after $t\approx100T_0$.

\subsection{The torque acting on the planet} 
Figure \ref{fig:Tdt} shows that the torque on the planets with $1M_{\oplus}\leq M_{p}\leq 5M_{\oplus}$ exhibits a clear periodic pattern 
since $t=100T_{0}$. This pattern stretches as time goes by. For planets with $M_{p}\geq 7M_{\oplus}$, at some point,
it changes and initiates a new different pattern. The pattern transition occurs approximately at $460T_{0}$ for $M_{p}=7M_{\oplus}$,
at $200T_{0}$ for $M_{p}=10M_{\oplus}$, and at $110T_{0}$ for $M_{p}=15M_{\oplus}$. 
Inspection of Figure \ref{fig:semi} indicates
that these transition times roughly coincide with the times at which planets stop migrating.
For $M_{p}=15M_{\oplus}$, the planet migrates inward so fast to the stalling radius 
that the first pattern appears only twice.

Figure \ref{fig:radial_torq} shows the radial torque distribution, $d\Gamma/dr$, for planetary masses of $1M_\oplus$ and $3M_\oplus$ at different times. 
The shape of the profile is different to the profile in the classical problem of a planet in a disc. In the latter, $d\Gamma/dr$
is antisymmetric with respect to the location of the planet. Here, the vortex formation causes that the maximum and minimum values of the radial torque distribution occur very close to the edges of the gap, which are located at $r_\mathrm{ie}=0.75r_0$ and $r_\mathrm{oe}=1.25r_0$, respectively.

We emphasize that the peak values of the radial distribution of the torque occur in time periods similar to the time it takes for the vortex to align azimuthally with the planet. For instance, in the case of $1M_\oplus$ at $t=500 T_{0}$, 
$d\Gamma/dr$ has not yet reached its maximum amplitude and in turn the main vortex is
at about 70 degrees behind the planet. On the other hand, at $t=502 T_{0}$, the radial 
distribution of the torque reaches its maximum value, which occurs when the main vortex
and the planet are close to aligning in azimuth. This behaviour is also reflected in the
total torque; its maximum value occurs when the planet is almost at its closest distance
to the main vortice, as can be observed in the cases of $M_p=1M_\oplus$, $M_p=7M_\oplus$ and $M_p=15M_\oplus$ at $t=500T_0$ (see red dashed lines in Fig. \ref{fig:Tdt} and the respective density maps in Fig. \ref{fig:DV-maps}).

\section{Discussion}
 \label{sec:discussion}
\subsection{A simple model for the torques acting on the planet in the migrating phase} \label{sec:simple_model}

In order to gain physical insight on the origin of the temporal behaviour of the
torque, on the migration rate, and on the stalling radius, we consider a simplified semi-analytical 
2D model. 

The total specific torque exerted on the planet has three components
\begin{equation}
\Gamma_{T}= \Gamma_{\rm L} + \Gamma_{\rm CR} + \Gamma_{V},
\end{equation}
where $\Gamma_{\rm L}$ is the Lindblad torque, $\Gamma_{\rm CR}$ the 
corotation torque and $\Gamma_{V}$ is the torque arising from the underdense vortices. 
For the Lindblad torque, we will use the formula
derived by Tanaka et al. (2002) for 3D isothermal discs:
\begin{equation}
\Gamma_{\rm L}/\Gamma_{0} = -2.5 + 0.1\sigma,
\label{eq:tanaka}
\end{equation}
where $\Gamma_0$ is a reference torque given as
\begin{equation}
\Gamma_{0} \equiv \frac{q^{2}}{h_{p}^{2}} \Sigma_{p} r_{p}^{4} \Omega_{p}^{2},
\end{equation}
with $\Omega_{p}$ is angular frequency of the planet, $h_{p}$ the disc aspect ratio at $r_{p}$,
and $q\equiv M_p/M_\star$ is the planet-to-star mass ratio.
The index $\sigma$ as a function of $r$ for the initial surface density radial profile used
in our simulations is shown in the upper panel of Figure \ref{fig:CR_torque}.

For the corotation torque we use the unsaturated value. 
The validity of this assumption is discussed in Section \ref{subsec:hsa}. 
The unsaturated corotation torque is given by
\begin{equation}
\frac{\Gamma_{\rm CR}}{\Gamma_{0}} = \frac{3}{4} \left(\frac{3}{2}-\sigma\right) \left(\frac{x_{s}}{r_{p}}\right)^{4}\frac{h_{p}^{2}}{q^{2}},
\label{eq:corotation_tq_formula}
\end{equation}
where $x_{s}$ is the half-width of the horseshoe region (e.g., Ward 1991; Paardekooper et al. 2010). 
The bottom panel of Figure \ref{fig:CR_torque} shows the
specific torque $\Gamma_{\rm CR}/M_{p}$, together with $\Gamma_{\rm L}/M_{p}$ in this model, as a 
function of the orbital radius of the planet $r_{p}$. Note that the corotation torque is positive
for orbital radius larger than $1$. We warn that the expressions
for the torques are only valid for masses smaller than the thermal mass.

Finally, the torque component $\Gamma_{V}$ arises from the two underdense banana-shaped regions, 
representing the low-density vortices, rotating with different angular velocity 
around the central star. In this analytical approach, we will assume that
the underdense regions themselves do not feel differential rotation; they preserve their shape with time.
We denote $\delta \Sigma_{V}$ to the decrement in the surface density of the disc associated to the vortices, i.e.
after substracting the axisymmetric part (which does not contribute to the torque).
Figure \ref{fig:Sigma_simple_model} shows $\delta \Sigma_{V}(r,\phi)$ in this model
at two different times. We have assumed that the inner and outer underdensities rotate in the same
direction that the planet, with constant angular frequencies 
$\Omega_{\mathrm{in}}=1.20$ and $\Omega_{\mathrm{out}}=0.78$, respectively. 
For illustration, if we assume that the planet is located at $r_{p}=1.236$,
the orbital frequencies in the frame corotating with the planet (synodic frequencies)
are $\tilde{\Omega}_{\mathrm{in}}=0.47$ and $\tilde{\Omega}_{\mathrm{out}}=0.052$, respectively.

The torque $\Gamma_{V}$ acting on the planet is given by
$\Gamma_{V}=\mathbf{\hat{z}}\cdot (\mathbf{r}_{p}\times \mathbf{F}_{V})$ with
\begin{equation}
\mathbf{F}_{V}= \sum_{i=1}^{N_{\rm cells}} \frac{GM_{p}\,\delta\Sigma_{V,i} \,r_{i}\Delta r \Delta\phi} {\left[(\mathbf{r}_{i}-
\mathbf{r}_{p})^{2}+\epsilon^{2}\right]^{3/2}}\,(\mathbf{r}_{i}-\mathbf{r}_{p}),
\end{equation}
where $\Delta r$ and $\Delta \phi$ are the grid spacing in the radial and
azimuthal direction and $\epsilon=0.6H(r_{p})$ is the smoothing length in our 2D semi-analytical model.
Note that the specific torque $\Gamma_{V}/M_{p}$ does not depend on $M_{p}$. As $\delta\Sigma_{V}$
is more negative at the core of the vortex, the force acts to increase
the planet's angular momentum when the planet is in front of a vortex.

The evolution of the semi-major axis of the planet is given by
\begin{equation}
    \frac{dr_{p}}{dt}=\frac{2 v F_{\parallel} }{\Omega^{2}_{p} r_{p}M_{p}},
\end{equation}
where $v$ is the velocity of the planet and $F_{\parallel}$ is the disc force tangential to the orbit \citep[e.g.,][]{Burns1976,MD1999}. Since the planetary eccentricities remain small, we will approximate
$v\simeq \Omega_{p} r_{p}$ and $F_{\parallel}\simeq \Gamma_{T}/r_{p}$.

When the planet entries to the edge of the gap, the local gradient of the disc surface density increases 
($\sigma$ decreases) and the positive coorbital torque becomes larger \cite[see Eq. (\ref{eq:corotation_tq_formula});][]{Masset_etal2006,Roma2019}.
As a consequence, the migration rate of the planet decreases. 
In our simplified model, we can estimate the stalling radius. We first note that as the vortex
structures are assumed to maintain their shapes, the average of $\Gamma_{V}$ over synodic periods
is zero. Consequently, the torque $\Gamma_{V}$ does not contribute to a net radial migration. Thus, planet
migration stalls when $\Gamma_{\rm L}+\Gamma_{\rm CR}=0$. Assuming that $x_{s}\simeq 1.1\sqrt{q/h}$, it 
implies $\sigma = -0.864$, regardless of the mass of the planet. 
For the unperturbed surface density in our simulations, an index of $-0.864$ occurs at $r=1.205$.

Figures \ref{fig:tq_rp_simple_model_1Moplus} and \ref{fig:tq_rp_simple_model_5Moplus} illustrate how the 
different components of the torque vary on time in our semi-analytical model, for a planet initially at
$r_{p}=1.25$, with $M_{p}=1M_{\oplus}$ 
(Figure \ref{fig:tq_rp_simple_model_1Moplus}) and $M_{p}=5M_{\oplus}$ (Figure \ref{fig:tq_rp_simple_model_5Moplus}).
The temporal evolution of the semi-major axis is also shown. We took $\Omega_{\textcolor{magenta}{\mathrm{in}}}=1.20$ and $\Omega_{\textcolor{magenta}{\mathrm{out}}}=0.78$.

We see that $\Gamma_{V}$ has the same shape as obtained in the simulations. $\Gamma_{V}$ follows a pattern that is repeated every 
$1/\tilde{\Omega}_{\textcolor{magenta}{\mathrm{out}}}$ (in units of $T_{0}$). Note that the synodic frequency $\tilde{\Omega}_{\textcolor{magenta}{\mathrm{out}}}$
increases with time as the planet migrates inward. $\Gamma_{V}$ also shows large-frequency oscillations of small
amplitude, which have a period of $1/\tilde{\Omega}_{\textcolor{magenta}{\mathrm{in}}}$ (in units of $T_{0}$), and are produced by the interaction
with inner vortex.

The model reproduces the general features found in the simulations: the amplitude of $\Gamma_{T}$, the shape of $\Gamma_{T}$,
the migration rate, and the radial incursions of the planet. 
Note that the amplitude of $\Gamma_{T}$ is dominated by $\Gamma_{V}$ because it is much larger than the amplitude of 
$\Gamma_{\rm CR}$ and $\Gamma_{\rm L}$.

The amplitude of the specific torque $\Gamma_{V}/M_{p}$ hardly changes when we vary the mass of the planet. 
However, the migration rate does depend on the mass of the planet, because the radial migration is driven by 
$(\Gamma_{\rm L}+\Gamma_{\rm CR})/M_{p}$. In fact, the $5M_{\oplus}$ planet has almost reached the stalling
radius after $900$ orbits. As can be seen in the middle panel of Figure \ref{fig:tq_rp_simple_model_5Moplus}, 
$\Gamma_{\rm CR}+\Gamma_{\rm L}$ approaches to zero (with some oscillations) at later times.

The rapid switch in the temporal pattern of the torque seen in Fig. \ref{fig:Tdt} for $M_{p}\geq 7M_{\oplus}$,
when the planet is close to the stalling radius, is because the planet-vortex interaction
is able to break up the outer vortex forming two vortices.
The planet with $M_{p}=1M_{\oplus}$ at $t=500T_{0}$ is still migrating towards the gap 
and it is still unable to break up the outer vortex.
In the map for the planet with $M_{p}=1M_{\oplus}$ in Fig. \ref{fig:DV-maps}, only two vortices are visible
in the disc. 
However, the planets with $M_{p}=7M_{\oplus}$ and $15M_{\oplus}$ have already
halted their radial migration, and, in fact, three vortices are present in the disc.
These vortices survive until the end of the simulations. The innermost vortex has the largest
orbital frequency about the central star and drives the small peaks in the torque.
The two outer vortices rotate around the star at similar orbital frequencies.

To summarize this section, we find that this simple model can qualitatively account for the general behaviour of the
torque and planetary orbital radius observed in the simulations, as far as the backreation of the planet on the vortices is not important.
However, we note that as the planet approaches to the vortices (especially to the outer vortex), 
the vortices cannot be treated as rigid low-density structures rotating at constant frequency around
the central star. 
As shown in Figures  \ref{fig:DV-maps} and \ref{fig:streams}, the planets can even pass through the vortices. In the next sections we discuss different aspects of the vortex-planet interaction when the planets are close to the vortices. 

\subsection{Vortex-planet interaction: Comparison with previous work}
\citet{Masset_etal2006} studied the migration of low-mass planets in the edge of a cavity (or surface density transition),
and showed that the migration is halted due to the positive contribution of the corotation torque.
Damping radial incursions of the planet (modulations) were also observed when the planet is close to the trapping 
radius due to the interaction with the anticylonic vortex formed at the top of the pressure bump.

\citet{Ataiee_etal2014} studied the gravitational interaction of an anti-cyclonic vortex with a planet.
They found that the planet is captured in a triangular equilibrium point in the three body system
formed by the star, the planet and the vortex. This capture is possible because anti-cyclonic vortices
have a larger density as behave as blob of matter. In our case, the vortices are underdense
and produce a repulsive effect on the planet in the azimuthal direction.
It is easy to show that there is no triangular equilibrium points in the circular restricted 
three body problem composed by the central star, the planet and the cyclonic vortex. If trapped
in a corotation resonance, the only possibility is that the planet lies in a collinear Lagrangian point. 
Whereas this is not the case in the simulations described in Section \ref{sec:results},
we show a case where the planet and the vortex are collinear with the central star in
the Appendix.

Very small, nearly unnoticeable, radial modulations are observed in the experiments of \citet{Ataiee_etal2014}. The reason is that being the vortex at the top of a dense ring, 
the Lindblad and corotation torques lead the planet to rapidly migrate inwards and to become locked in corotation with the vortex. Thus, the interval of time at which the 
planet can be pushed to an outer orbital radius becomes very short. In the experiments of Fig. \ref{fig:semi}, 
the radial incursions are much
larger because the planet stops its migration at a radius slightly larger than the orbital 
radius of the core of the outer vortex. In the Appendix, we show a case where the radial incursions damp with time because
the planet is swallowed by the vortex.

\begin{figure}
\includegraphics[width=0.5\textwidth]{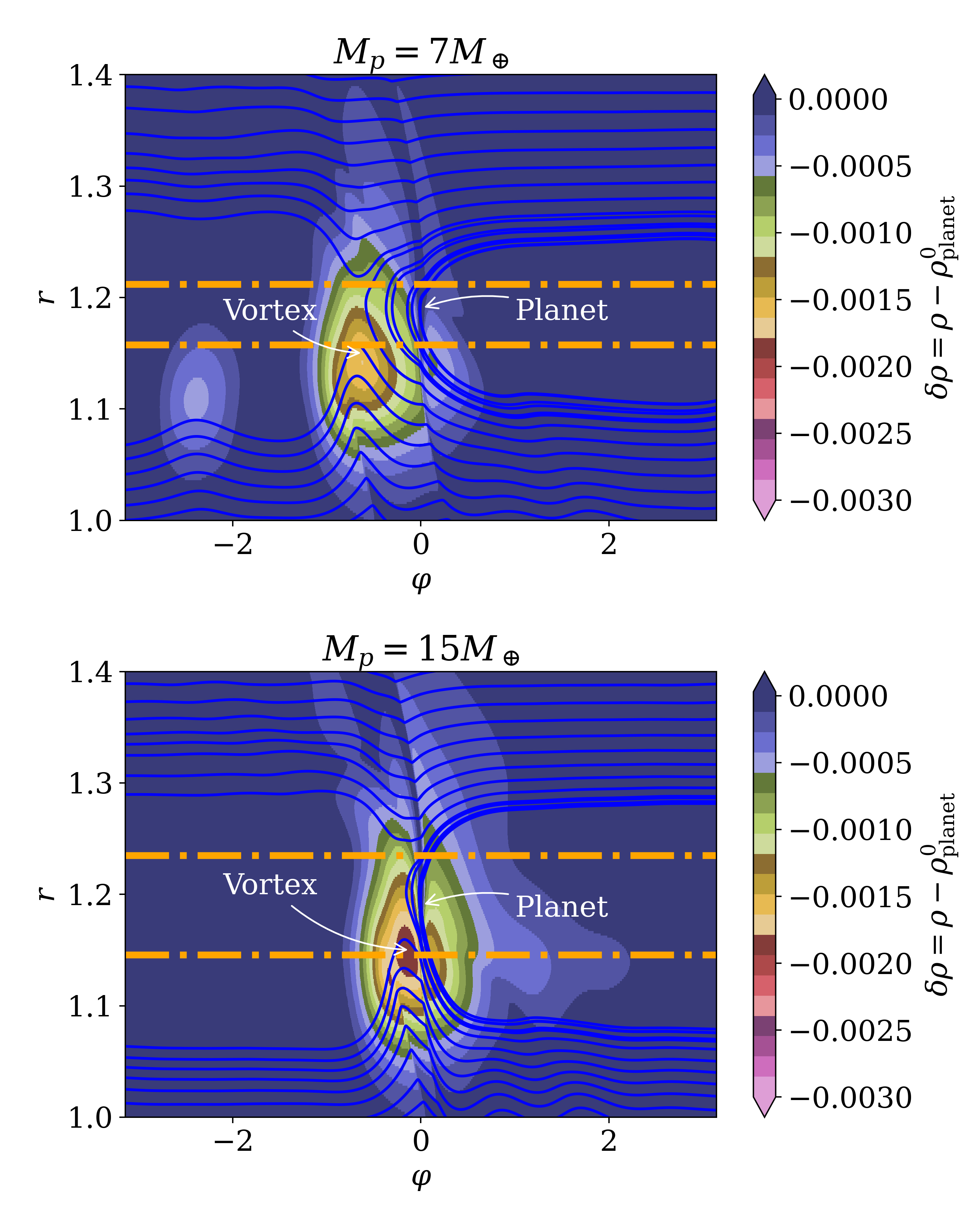}
  \caption{\textbf{Midplane} perturbed density, $\delta\rho=\rho-\rho_\mathrm{planet}^0$, around the co-rotation region for planetary masses of $M_p=7 M_\oplus$ (\textit{top panel}) and $M_p=15 M_\oplus$ (\textit{bottom panel}), at $t=500 T_{0}$ after inserting the planet. Note that the gas streamlines (solid blue lines) execute a U-turn in the azimuthal position where the underdense vortex is located. The dot-dashed lines show the half-width of the horseshoe region predicted by Eq. (\ref{eq:xs}).}
\label{fig:streams}
\end{figure}

\begin{figure}
\includegraphics[width=0.5\textwidth]{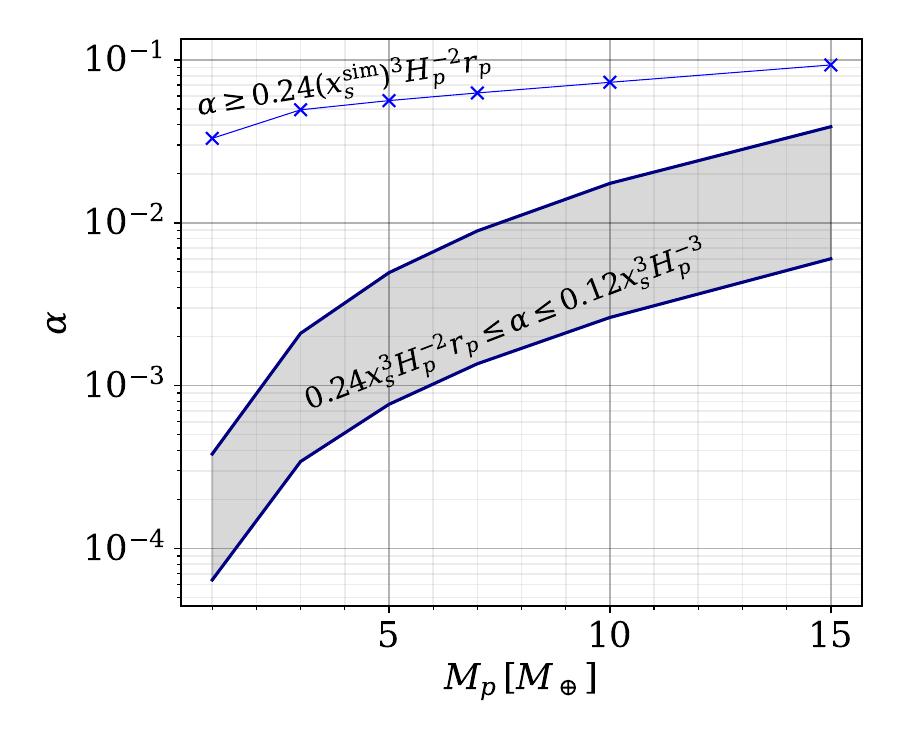}
  \caption{Viscosity domain according to Eq. (\ref{eq:ineqNew}) as a function of $M_{p}$, using the horseshoe half-width $x_s$ given
  in Eq. (\ref{eq:xs}) (shaded region). The lower bound given in Eq. (\ref{eq:ineqNew}) but now using $x_s^\mathrm{sim}$ at $t=500  T_{0}$
  is also shown (blue curve with crosses).}
\label{fig:alpha_i}
\end{figure}

\subsection{Horseshoe region analysis}
\label{subsec:hsa}

The libration timescale in which a fluid element orbiting at radius $r_e=r_p(1+x_s/r_p)$ execute two orbits in the frame corotating with the planet, is given as \citep[see][]{M2001,M2002}:
\begin{equation}
    \tau_\mathrm{lib}=\frac{4r_p}{3x_s}\frac{2\pi}{\Omega_p},
    \label{eq:Tau_hs}
\end{equation}
if this timescale is smaller than the timescale of the viscous evolution of the disc,
\begin{equation}
\tau_\mathrm{vis}\approx\frac{x_s^2}{\nu},
    \label{eq:Tau_nu}
\end{equation}
then the corotation torque will stay unsaturated. If the latter occurs, the fluid elements in the librating region are pushed into the horseshoe region. 

Other possible mechanisms that may give rise to new fluid elements within the horseshoe region are the migration of the planet itself \citep{Paar_etal2010,Paar2014,Roma2019}, and the disturbances generated by a vortex and the spiral waves that it emits \citep{Chametla_Chrenko2022,Chame2023}.

For the horseshoe torque to remain unsaturated, the following inequality 

\begin{equation}
\tau_\mathrm{U-turn}\leq\tau_\mathrm{visc}\leq\frac{\tau_\mathrm{lib}}{2}
    \label{eq:ineq}
\end{equation}
should be satisfied \citep{BaruP2010,BM2013}. Here,  $\tau_\mathrm{U-turn}\approx h_p\tau_\mathrm{lib}$
is the horseshoe U-turn time. Eq. (\ref{eq:ineq}) may be cast as
\begin{equation}
0.24x_s^3H_p^{-2}r_p\leq\alpha\leq0.12x_s^3H_p^{-3},
    \label{eq:ineqNew}
\end{equation}
where $H_p$ is the pressure scale of the disc calculated at $r=r_p$. Remarkably, in all our simulations, the outer vortex yields to a widening of the horseshoe region (see, for instance, Fig. \ref{fig:streams}). This implies that the half-width of horseshoe region $x_s$ may differ from the value inferred in the absence of any vortex:
\begin{equation}
x_s=\frac{1.05(q/h)^{1/2}+3.4q^{7/3}/h^6}{1+2q^2/h^6}r_p
    \label{eq:xs}
\end{equation}
\citep{JM2017}.

Note that the width of the horseshoe region does not depend substantially on the surface density profile \citep{Masset_etal2006}.
In our simulations, the widening of the horseshoe region is mainly caused by the low pressure generated in the cyclonic vortex. The low pressure region inside the vortex produces
an asymmetrical distortion of the streamline pattern close to the planet and an azimuthal shift of the stagnation points (see Fig. \ref{fig:streams}). The widening of the horseshoe region, $S\equiv x_{s}^{\mathrm{sim}}/x_{s}$, is not constant for the different planets, but it is a function of the mass of the planet.
When the vortex and the planet are aligned azimuthally (that is, null angular separation) or trapped in corotation resonance (as in the case shown in the Appendix),
we find that 
\begin{equation}
S(q)=\frac{3.4h_0}{\sqrt{q+q_0}}
    \label{eq:fit}
\end{equation}
where $q_0=0.5M_\oplus/M_\star$. 
For instance, for a planetary masses of $M_p=1M_\oplus$ and $M_p=15M_\oplus$, we find that $S=8$ and $2.5$, respectively.

Fig. \ref{fig:alpha_i} shows the range of $\alpha$ that satisfies the inequality (\ref{eq:ineqNew}) for
different values of $M_{p}$, using $x_{s}$ given in Eq. (\ref{eq:xs}). For $M_{p}=7M_{\oplus}$, we see that 
range of $\alpha$ to keep the corotation torque unsaturated spans from $10^{-3}$ to $10^{-2}$. Since these value of $\alpha$ 
are similar to the values in the gap of our simulations, the corotation torque would remain unsaturated if $x_{s}$ were given
by Eq. (\ref{eq:xs}). However, if instead of $x_{s}$ as given by Eq. (\ref{eq:xs}), we use $x_{s}^{\mathrm{sim}}$, 
the viscosity needed to keep the corotation torque unsaturated should be larger than
$\alpha=3.3\times 10^{-2}$ for $M_{p}=1M_{\oplus}$ and larger than $\alpha=9.3\times 10^{-2}$ for $M_{p}=15M_{\oplus}$ (see curve with crosses in Fig.
\ref{fig:alpha_i}). However, the viscosity in our simulations is below these values. Since the corotation should not be saturated in order to balance the differential
Lindblad torque, we argue that the corotation torque remains unsaturated (or partially unsaturated) due to the action of the cyclonic vortex 
in the horseshoe region of the planet (see Fig. \ref{fig:streams}). 

\begin{figure}
\includegraphics[width=0.5\textwidth]{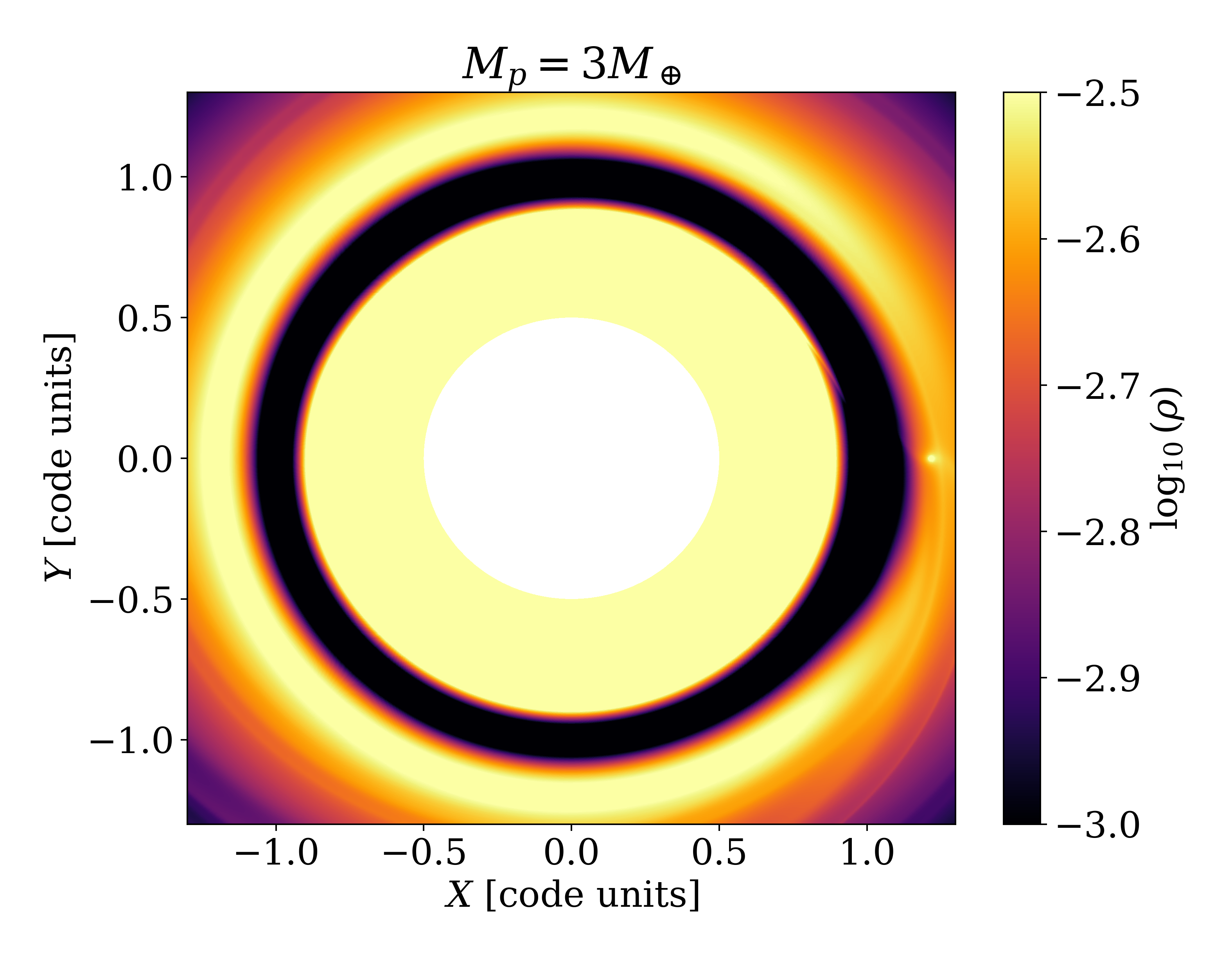}
  \caption{Asymmetric gas density distribution in the gap region produced by a cyclonic vortex when a low-mass planet of $M_p=3M_\oplus$ is embedded in the disc, at $t=500 T_{0}$.}
\label{fig:gapasim}
\end{figure}

\begin{figure}
\includegraphics[width=0.5\textwidth]{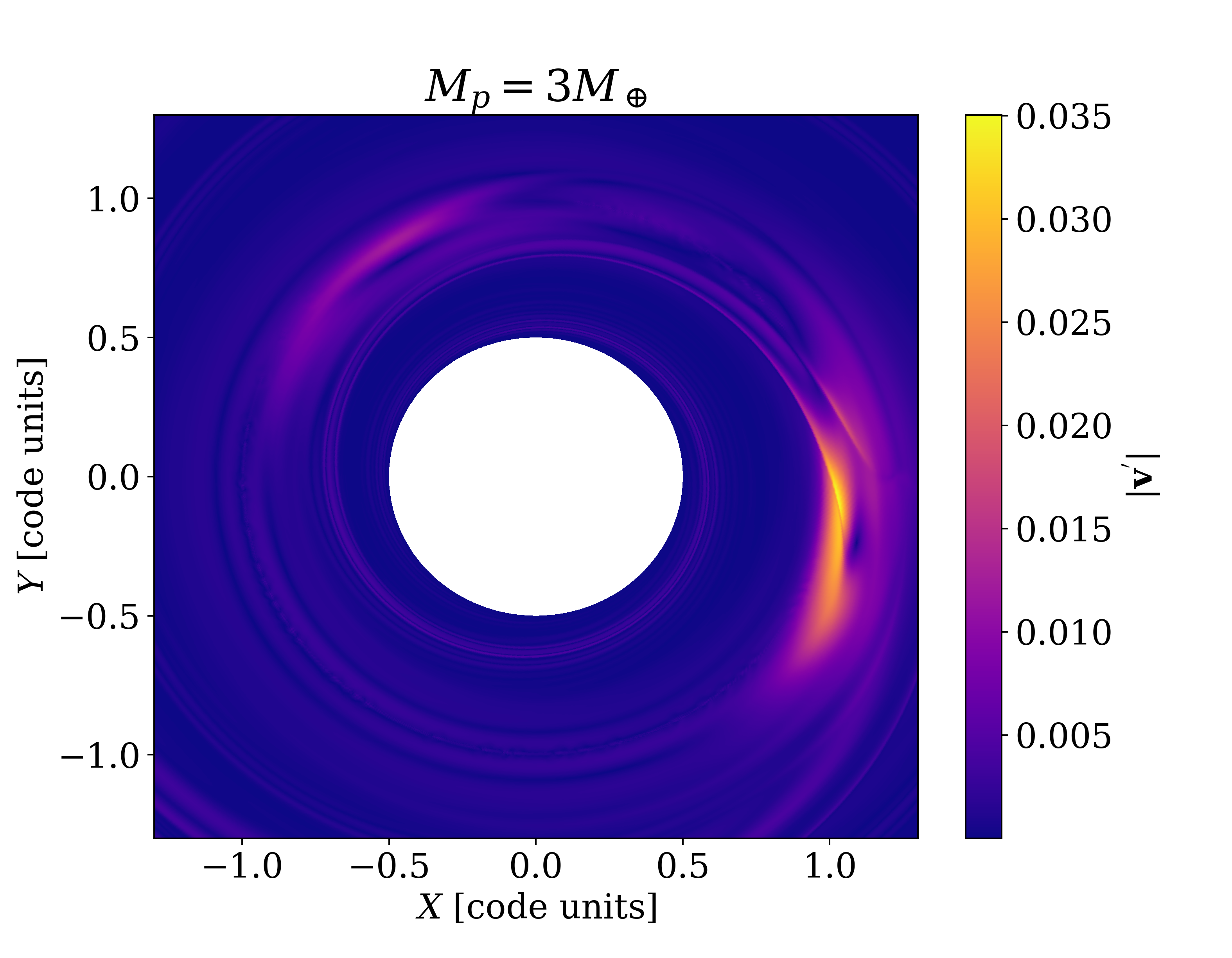}
  \caption{Velocity field perturbations $\abs{\mathbf{v}'}\equiv[(v_r-\langle v_r\rangle)^2+(v_\theta-\langle v_\theta\rangle)^2+(v_\phi-\langle v_\phi\rangle)^2]^{1/2}$, in the gap region produced by a cyclonic vortex when a low-mass planet of $M_p=3M_\oplus$ is embedded in the disc, at $t=500T_{0}$.}
\label{fig:VRes}
\end{figure}

\subsection{Effects on planet formation}

Some of the potential effects of the planet trapping scenario presented here are the following. 

The radial excursions of the planet due to the interaction with vortices in the gap, as illustrated in Fig. \ref{fig:semi}, can lead to an expansion of the so-called feeding zone for the migrating planet, the region within the protoplanetary disc from where solids can be gravitationally captured by the planet. Typically it is considered to include the area within a few Hill radii of the semi-major axis of the planet \citep{Armitage_2009}. An estimation of this effect for the planet masses under consideration, 1 - 15 $M_\oplus$, around a 1 $M_\odot$ star, indicates that the feeding zone width, and hence the final planet mass, is generally less than 50\% greater than the normal value. In addition, our results seem to indicate that the amplitude of the radial excursions is not strongly dependent on the planet mass, so this is most probably not a runaway scenario.     

In the classical picture, anticyclonic vortices form at the pressure maxima at the edges 
of the gaps. It is thought that these anticyclonic vortices can trap efficiently solid particles \citep{Barge&Sommeria1995,Ataiee_etal2014}. On the contrary, cyclonic vortices are characterized by a central pressure minimum and, hence, do not trap solid particles. This implies that, if gaps in protoplanetary discs are inhabitated by vortices as those studied here, planet formation is not favoured in such regions, and dust would preferably by driven to the edges of the gap and/or dispersed azimuthally. In this sense, cyclonic vortices would operate as expelling eddies that could contribute to prevent the radial drift of
large dust particles. However, the enhancement
in the $\alpha$-viscosity parameter within the gap will increase the turbulent diffusion
of dust particles. As a result,  we expect the formation of wider dust rings in
gaps with cyclonic vortices than in the 
classical case of a pressure bump with anticyclonic vortices.

An additional interesting possibility, at the moment a speculation to be explored in future studies, is the consequence of trapping additional Earth and super-Earth planets as they migrate inward from radial locations further out in the disc. In the presence of a gaseous discs, it is possible that such planets can be trapped in mean motion resonance with the inner planet, as suggested in the scenario studied by \cite{Pierens&Nelson2008}.

\subsection{Observational implications}

Since the vortices formed at the edges of the gap are underdense and cyclonic (see Fig. \ref{fig:DV-maps}), they can prevent dust accumulation, so it can be very difficult to observe them through standard techniques. However, effects that these produce can be identified, since they generate asymmetries at the edges of the gap that evolve over time. Therefore, the vortices reported in this study can be considered as candidates to explain the different asymmetries observed in the gaps of protoplanetary discs \citep[][]{Andrews2018,Isella2019,Step2023}.

Another interesting consequence of these vortices is that they could be used as tracers of the stagnation orbital radius of the planets. As can be seen in Fig. \ref{fig:DV-maps}, the planet stops radial migration at the same radius as the orbital radius of the vortex. In fact, 
the planet passes through the vortex (being trapped inside or not) where the asymmetry in the gap occurs (see for instance Fig. \ref{fig:gapasim}). We expect to have low-mass planets
in regions near the edge of the gap. This can help observationally search for low-mass planets.
Since these vortices should have a strong impact on the local kinematics of the gas (see Fig. \ref{fig:VRes}), it should be possible to detect and identified them through the analysis of perturbations in the velocity field, similar to those used in the indirect detection of massive accreting planets in gaps \citep[see][for a review]{Pinte2023}.

So far there are only very few massive planets detected observationally \citep[e.g.,][]{ Benisty2021}.
A search for low-mass planets can be even more challenging. Our results that low-mass planets can generate cyclonic vortices give an important indirect possibility of detecting them in protoplanetary discs that exhibit asymmetric gaps.

\section{Conclusions} \label{sec:conclusions}

We have performed 3D hydrodynamical simulations of the migration of
low-mass planets embedded in a globally-isothermal disc with a gap in
its surface density.  The gap is sustained by a viscosity bump. When
the planet is introduced into the disc, it excites density waves
that promote the formation of underdense cyclonic vortices at the edges
of sufficiently deep gaps.
While the increasing density gradient in the outer edge of the gap should 
yield to an enhancement of the positive corotation torque that counteracts 
inward migration, the vortices formed in the gap modifies the flow structure  
in the coorbital region of the planet and, therefore, alter the corotation
torque.  Our main aim has been to investigate the interaction of the planet 
with the cyclonic vortices as the planet migrates towards the gap.

Initially, when the planet is far away from the gap, the vortices can be treated
as rigid underdense entities. In this stage, the total torque
acting on the planet varies periodically over the mutual (vortex and planet) 
synodic period, and its amplitude is dominated by the gravitational interaction with the
main vortex. Afterwards, the overall inward migration of the planet stops 
at the outer edge of the gap. However, in some cases, the planet executes 
radial incursions even in the trapped state, due to the ``repulsive'' 
interaction with the underdense main vortex. 
In other cases, the planet and the vortex are locked in corotation and,
thus, the flow reaches a steady state in the frame corotating with the
planet. In this steady state, the flow around the planet presents
a front-back asymmetry, which is responsible for a positive torque on
the planet that counteracts the negative Lindblad torque.

The formation of this type of underdense cyclonic vortices can have important observational implications since it can generate asymmetries in the 
gas density within the gap that could explain the similar asymmetries that have been observed in different protoplanetary discs \citep[see][]{Andrews2018}.

\section*{Acknowledgements}
We are grateful to the referee for her/his constructive and careful report. The work of R.O.C. was supported by the Czech Science Foundation (grant 21-23067M). Computational resources were available thanks to the Ministry of Education, Youth and Sports of the Czech Republic through the e-INFRA CZ (ID:90254). C.C.-G. acknowledges support from UNAM DGAPA-PAPIIT grant IG101224 and from CONACyT Ciencia de Frontera project ID 86372. The work of O.C. was supported by the Charles University Research program (No. UNCE/SCI/023).

\section*{Data Availability}

 The FARGO3D code is available from \href{https://fargo3d.github.io/documentation}{https://fargo3d.github.io/documentation}. The input files for generating our 3D hydrodynamical simulations will be shared on reasonable request to the corresponding author.



\bibliographystyle{mnras}
\bibliography{Manuscript} 




\appendix
\section{Planets trapped in shallower gaps}
\label{ap:appendixA}

\begin{figure}
\includegraphics[width=0.5\textwidth]{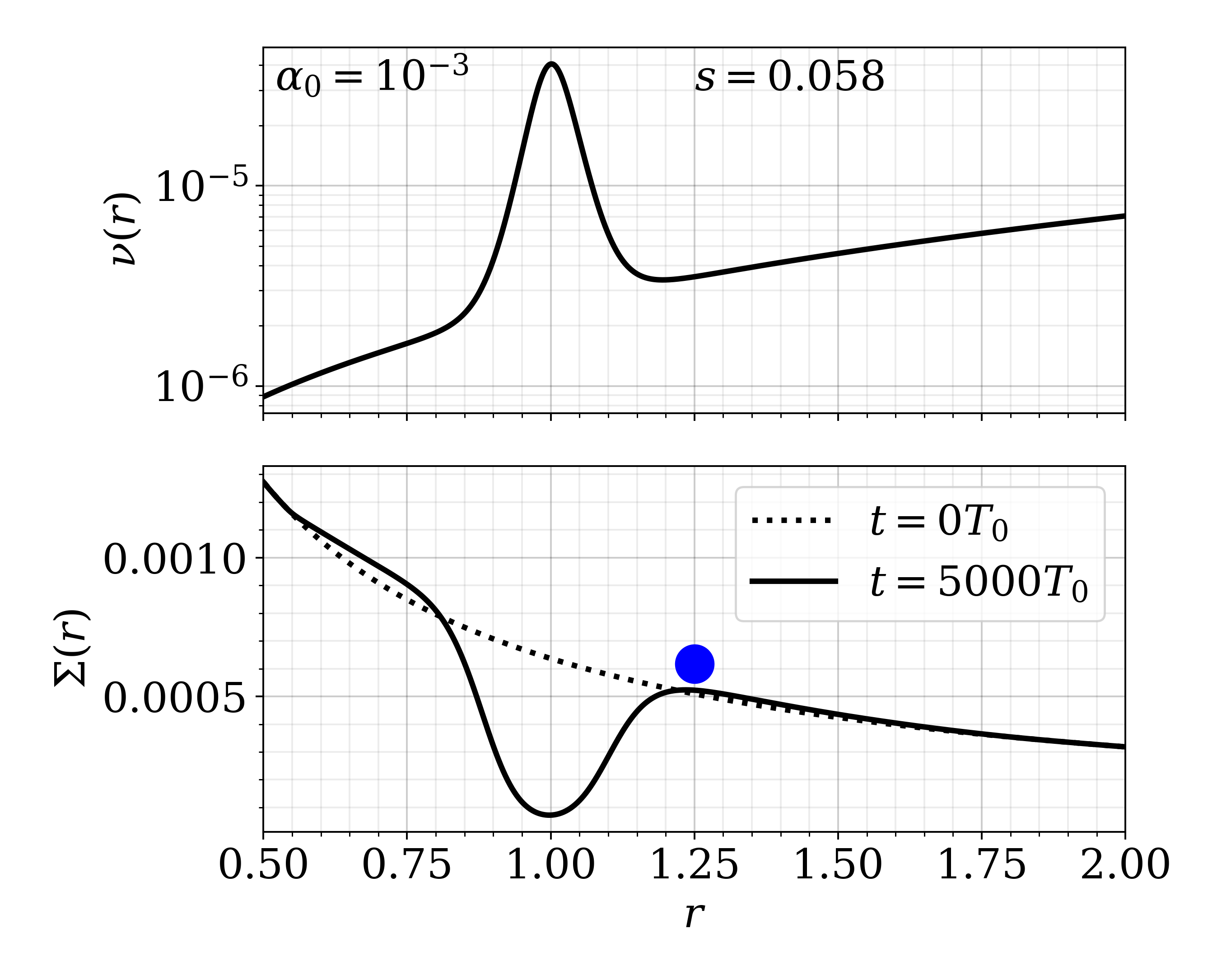}
 \caption{\textit{Top:} Radial profile of the kinematic viscosity $\nu$, parameterized by Eq. (\ref{eq:nu}). \textit{Bottom:} Radial profile of the surface density when the planet is introduced into the disc.}
\label{fig:sgap}
\end{figure}

\begin{figure*}
\includegraphics[width=19cm,height=14cm]{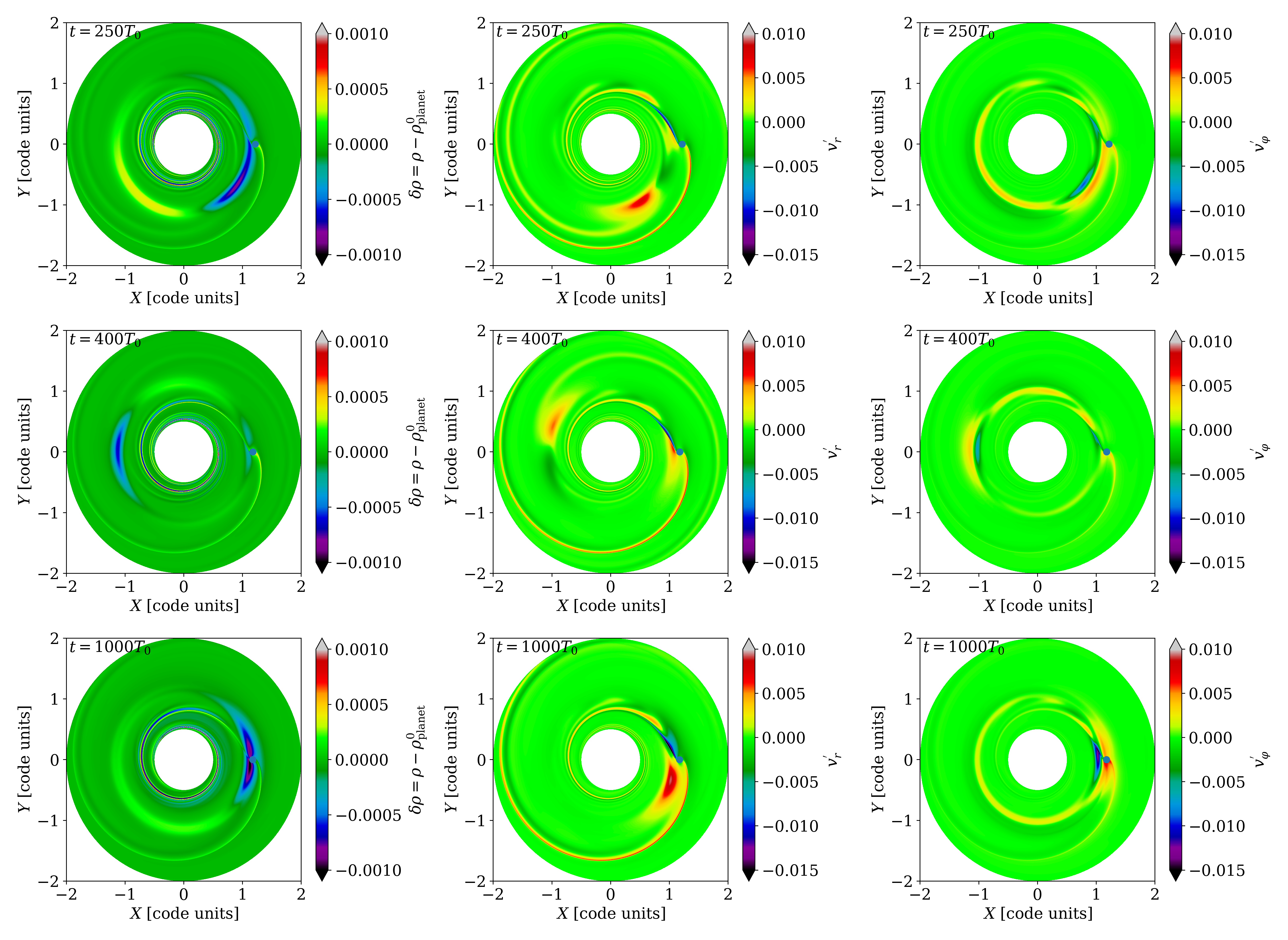}
  \caption{\textbf{Midplane} volume density perturbation $\delta\rho=\rho-\rho_\mathrm{planet}^0$ (left column),
radial velocity perturbations $v'_{r}=v_{r}-\left<v_{r}\right>$ (central column) and
azimuthal velocity perturbations $v'_{\varphi}=v_{\varphi}-\left<v_{\varphi}\right>$ (right column)
at $t=250T_{0}$ (upper row), $t=400T_{0}$ (middle row) and $t=1000 T_{0}$ (lower
row).
The planet has a mass of $7M_{\oplus}$ and rotates counter-clockwise.
The parameters of the viscosity profile are $s=0.058$, $r_{1}=0.9$ and $r_{2}=1.1$.}
\label{fig:h0058_33}
\end{figure*}

\begin{figure*}
\includegraphics[scale=0.315]{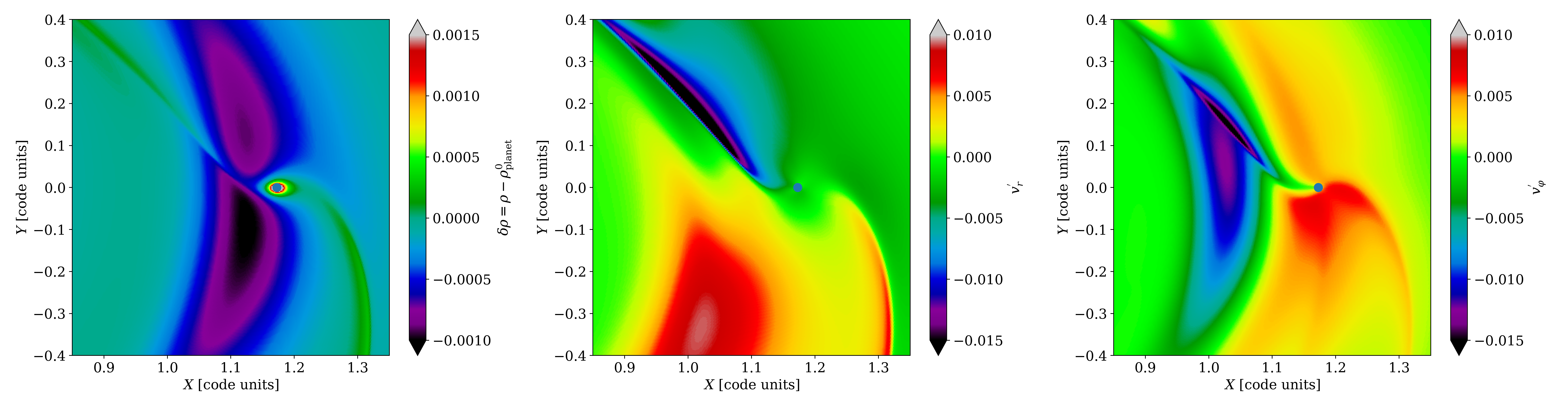}
  \caption{Zoom of the flow pattern around the planet
at $t=1000T_{0}$. \textit{Left:} Perturbations in density. \textit{Middle:} Radial
velocity perturbations. \textit{Right:} Azimuthal velocity perturbations. }
\label{fig:h0058_zoom}
\end{figure*}

\begin{figure}
\includegraphics[width=0.5\textwidth]{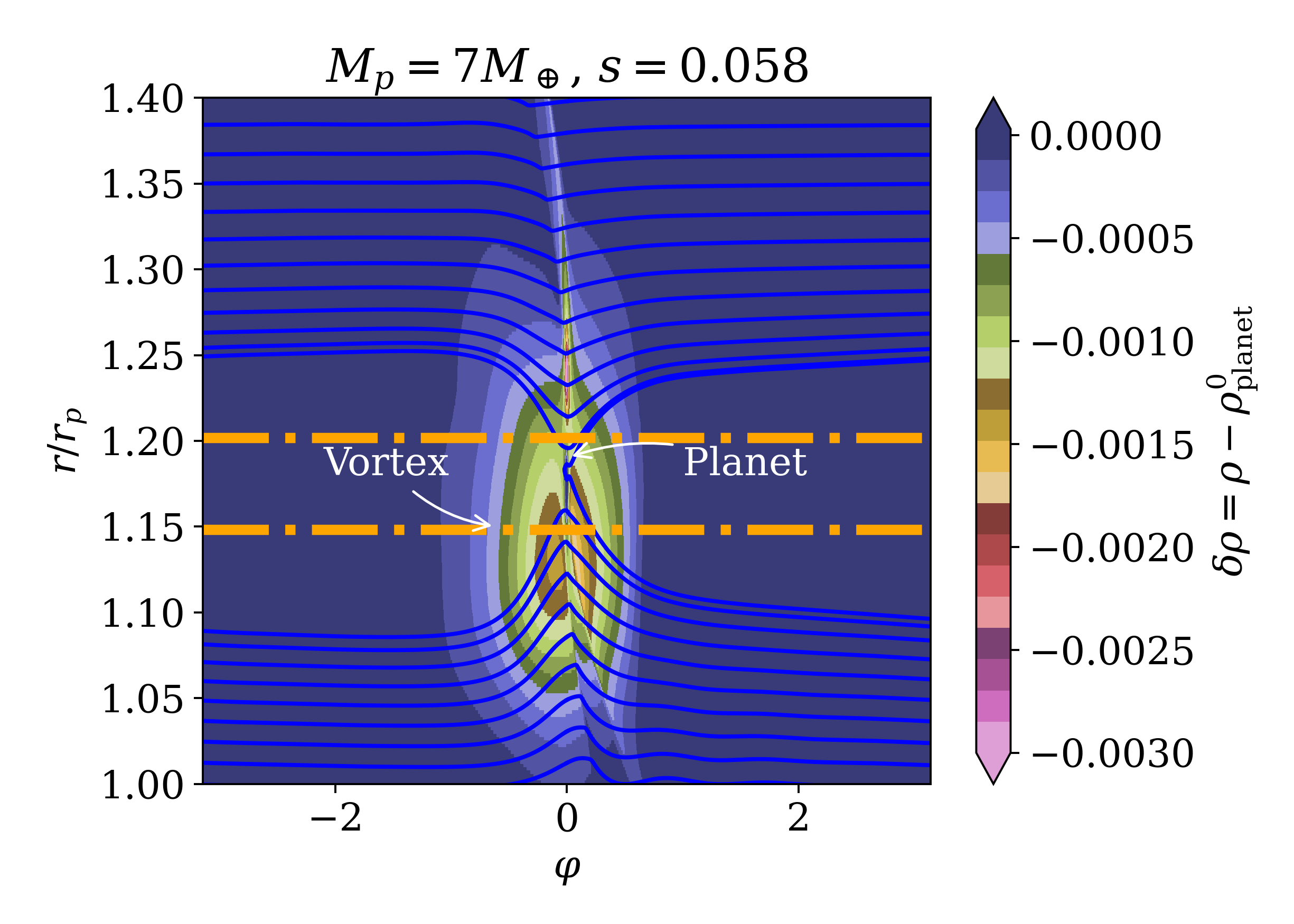}
  \caption{\textbf{Midplane} perturbed density, $\delta\rho=\rho-\rho_\mathrm{planet}^0$, around the co-rotation region for a planet of mass of $M_p=7 M_\oplus$. As in the previous cases, the gas streamlines (solid blue lines) execute a U-turn in the azimuthal position where the underdense vortex is located.}
\label{fig:shallower}
\end{figure}
\begin{figure}
\includegraphics[width=0.45\textwidth]{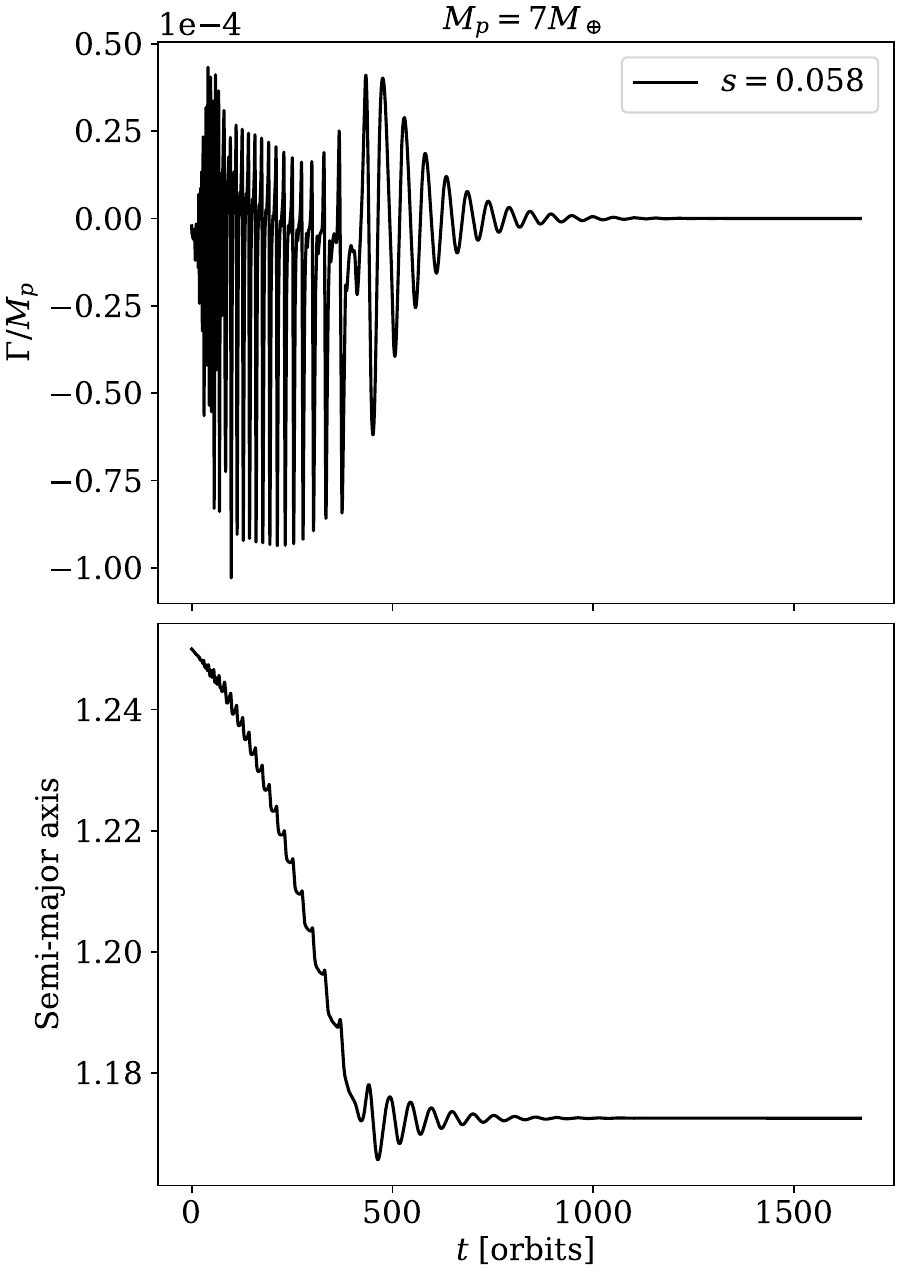}
  \caption{Temporal evolution of the specific torque $\Gamma/M_{p}$ on a planet of mass $M_{p}=7M_{\oplus}$
  (upper panel) and semi-major axis (lower panel), 
  in a model with $s=0.058$, $r_{1}=0.9$ and $r_{2}=1.1$.}
\label{fig:tq_semi_h0058}
\end{figure}

For the gap parameters considered in Sections \ref{sec:initial} and \ref{sec:results}, we have shown that the
planet is continuously interacting with the cyclonic vortices formed in the gap.
When the planet is in the migrating phase (far enough from the vortices), 
the vortices can be treated as ``rigid'' underdense structures. 
However, when the planet approaches to 
the orbital radius of the vortices,  the vortices-planet dynamics becomes very complex;
the vortices break up into smaller vortices and the planet can go through the outer
vortex repeatedly.

It is expected that the number, strength and circulation of the vortices depend on the 
depth and width of the gap. The probability of the formation of cyclonic vortices is
likely to be reduced as the gap is shallower. In order to explore if the formation of
cyclonic vortices is still feasible in a shallower gap, 
we have reduced the amplitude of the viscosity bump. In a shallower gap, we expect
less intense vortices and, therefore, 
a reduction in the strength of the interaction between the planet and the vortices.

We have adopted $s=0.058$ (we keep the same values of $r_{1}=0.9$ and 
$r_{2}=1.1$), implying that the viscosity at $r=1$ is a factor $4/7$ smaller then 
it is for $s=0.05$. Figure \ref{fig:sgap} shows the viscosity and surface
density profiles.

We now present the results when a planet of $7M_{\oplus}$ is inserted in the disc.
Figure \ref{fig:h0058_33} shows the density perturbations $\delta\rho=\rho-\rho_\mathrm{planet}^0$ and velocity perturbations $v_{r}'\equiv
v_{r}-\left<v_{r}\right>$ and
$v_{\varphi}'\equiv v_{\varphi}-\left<v_{\varphi}\right>$ in the equatorial plane, at three 
different times ($t=250 T_{0}, 400T_{0}$ and $1000T_{0}$). Here the brackets 
$\left<...\right>$ denote azimuthally averaged values. In the maps of the density
perturbations (left panels), the outer spiral wave excited by the planet plus 
two arc-shaped regions (one overdense and one underdense, which corresponds 
to a cyclonic vortex) are visible. The signature of the vortex is also seen as two-lobed
structures in the maps of the perturbations in radial velocity and
as two stripes in the maps of the azimuthal velocity perturbations. In the vortex,
the radial velocity perturbation changes sign along the azimuthal, whereas the
azimuthal velocity perturbation changes sign along the radial direction. 

At $t=250T_{0}$, the vortex is approaching to the planet, being the angular separation between the planet and the vortex of $\sim 40^{\circ}$. At $t=400 T_{0}$, even
though the vortex and the planet are in conjunction, the maps of the velocity
perturbations show a complex pattern around the planet. At $t=1000T_{0}$, 
the planet has been captured by the vortex and both corotates in a steady state.
A zoom of the flow around the planet is shown in Fig. \ref{fig:h0058_zoom}
and the streamlines in Fig. \ref{fig:shallower}.

\begin{figure*}
\includegraphics[scale=0.41]{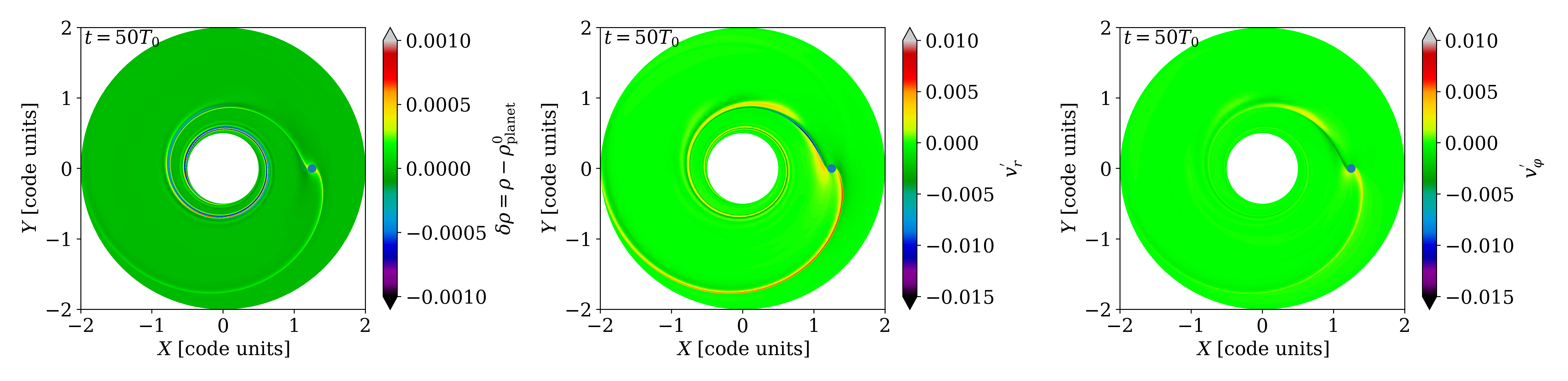}
  \caption{Volume density perturbation $\delta\rho$ (left panel),
radial velocity perturbations $v'_{r}$ (central panel) and
azimuthal velocity perturbations $v'_{\varphi}$ (right panel) at $z=0$ and $t=50T_{0}$. 
The parameters of the gap profile are $s=0.085$, $r_{1}=0.88$ and $r_{2}=1.12$.
The planet rotates counter-clockwise.}
\label{fig:h0085}
\end{figure*}

Figure \ref{fig:tq_semi_h0058} shows the total (specific) torque on the planet as a function of time.
Between $t=70T_{0}$ and $300T_{0}$, the amplitude of the total torque on the planet, $\Gamma$, 
is dominated by the
gravitational interaction with the underdense arc-shaped region plus the overdense region.
The tamdem effect of both contributions produce that $\Gamma/M_{p}$ oscillates 
between $\sim -9\times 10^{-5}$ and $\sim 2\times 10^{-5}$.
After $t=300T_{0}$, the arc-shaped overdense weakens.
At $400T_{0}$, $\Gamma$ undergoes a transition in its behaviour and a new phase, where the planet
will eventually be locked to the cyclonic vortice, starts. After $t=400T_{0}$, $\Gamma$
displays a classical damping pattern in time, until it becomes
strictly zero in the stationary state (see upper panel in Figure \ref{fig:tq_semi_h0058}). 

The lower panel of Fig. \ref{fig:tq_semi_h0058} shows the temporal evolution of the semi-major 
axis of the planet. We see that between $0$ and $400T_{0}$,
$d^{2}r_{p}/dt^{2}<0$ or, in other words, the migration rate increases
until it stops suddenly at $400T_{0}$. The stalling radius is $1.17r_0$, which
is slightly smaller than for $s=0.05$. After $400T_{0}$, the planet and the
vortex rotate in tandem, with a small damping librational motion (which are visible
as oscillations in the semi-major axis between $t=400T_{0}$ and $t=750T_{0}$). 
After $t=900T_{0}$, the libration is totally damped and the flow becomes stationary
in the frame rotating with the planet.
In this stationary configuration, the planet and the vortex are aligned
in opposition (see lower panels in Figure \ref{fig:h0058_33}). Fig. \ref{fig:h0058_zoom} and \ref{fig:shallower} clearly show that
the vortex modifies the flow around the planet, producing a front-back asymmetric,
which is responsible for a positive torque on the planet that balances the differential
Lindblad torque. The origin of this asymmetry  can be explained as arising from
the radial shift between the planetary radius (with $r_{p}=1.17r_{0}$) and the vortex
center (at $r=1.04r_{0}$; this corresponds to the radius at which $v_{\varphi}'$ changes sign).

Finally, we have carried out a simulation with $s=0.085$ and $r_{1}=0.88$
and $r_{2}=1.12$. In this case, the $\alpha$ viscosity parameter is a factor of $3$
smaller than it is in our reference model.
A snapshot of the density and velocity perturbations at $t=50T_{0}$ 
is shown in Figure \ref{fig:h0085}. No vortex is formed. The pattern velocity around the
planet is similar to the pattern in Figure \ref{fig:h0058_33} at $t=400T_{0}$. This indicates
that the flow around the planet at $t=400T_{0}$, when the vortex is in conjunction in the simulation
of Figure \ref{fig:h0058_33}, has time to relax
to the same configuration as if there was no vortex.



\bsp	
\label{lastpage}
\end{document}